\documentclass[a4paper,11pt,reqno]{amsart}
\usepackage{amsmath,amsthm,amssymb}
\usepackage{enumerate}
\usepackage{graphicx}
\newcommand{\dd}{\mathrm{d}}
\newcommand{\ee}{\mathrm{e}}
\newcommand{\ii}{\mathrm{i}}
\theoremstyle{plain}
\newtheorem{thm}{Theorem}
\newtheorem{lem}[thm]{Lemma}
\newtheorem{prop}[thm]{Proposition}
\newtheorem{cor}[thm]{Corollary}
\theoremstyle{definition}
\newtheorem*{Def*}{Definition}
\newtheorem*{rems*}{Remarks}
\newtheorem*{rem*}{Remark}
\providecommand{\C}[1]{\mathcal{#1}}
\providecommand{\D}[1]{\mathbb{#1}}

\providecommand{\abs}[1]{\lvert#1\rvert}
\providecommand{\accol}[1]{\lbrace#1\rbrace}
\providecommand{\croch}[1]{\lbrack#1\rbrack}
\providecommand{\norm}[1]{\lVert#1\rVert}
\providecommand{\scal}[1]{\langle#1\rangle}
\renewcommand{\Im}{\mathop{\rm Im}}
\DeclareMathOperator{\const}{const.}
\DeclareMathOperator{\diag}{diag}
\DeclareMathOperator{\dist}{dist}
\DeclareMathOperator{\expect}{\D{E}}  %expectation
\DeclareMathOperator{\expecttilde}{\widetilde{\D{E}}}
\DeclareMathOperator{\expecthat}{\widehat{\D{E}}}
\newcommand{\HS}{\textup{HS}}           %Hilbert-Schmidt
\DeclareMathOperator{\prob}{\D{P}}     %probability
\DeclareMathOperator{\supp}{supp}
\DeclareMathOperator{\tr}{tr}
\begin{document}
%-------------------------------------------------------------%
\title[Localization near fluctuation boundaries]{Localization near fluctuation
boundaries\\ via fractional moments and applications}
%-------------------------------------------------------------%
\author[A.~Boutet de Monvel]{Anne Boutet de Monvel$^1$}
\author[S.~Naboko]{Serguei Naboko$^{2,3}$}
\author[P.~Stollmann]{Peter Stollmann$^4$}
\author[G.~Stolz]{Gunter Stolz$^3$}
\address{$^1$ IMJ, case 7012, Universit\'e Paris 7,
               2 place Jussieu, 75251 Paris, France}
\address{$^2$ Department of Mathematical Physics, St.\ Petersburg
University, St.\ Petersburg, Russia}
\address{$^3$ Department of Mathematics,
           University of Alabama at Birmingham, Birmingham, AL 35294, USA}
\address{$^4$ Fakult\"at f\"ur Mathematik,
           Technische Universit\"at, 09107 Chemnitz, Germany}
%-------------------------------------------------------------%
\date{\footnotesize\today} % September 2, 2005
\maketitle
%-------------------------------------------------------------%
\begin{abstract}
We present a new, short, self-contained proof of localization
properties of multi-dimensional continuum random Schr\"odinger
operators in the fluctuation boundary regime. Our method is based
on the recent extension of the fractional moment method to
continuum models in \cite{AENSS}, but does not require the random
potential to satisfy a covering condition. Applications to random
surface potentials and potentials with random displacements are
included.
\end{abstract}
%-------------------------------------------------------------%
\maketitle
%-------------------------------------------------------------%
\section{Introduction}

%-------------------------------------------------------------%
\subsection{Motivation}

We are concerned here with proving localization properties of
multi-dimensional continuum random Schr\"odinger operators in
the fluctuation boundary regime.

Such results were first found via the method of multiscale
analysis, which had been developed in the 80s to handle lattice
models and was later extended to the continuum (for a rather
complete history and list of references on multiscale analysis see
\cite{Stollmann} and for some of the more recent developments
\cite{Germinet/Klein}).

Later, the fractional moment method was developed
\cite{Aizenman/Molchanov} as an alternative approach to the same
problem, also initially for lattice models. It leads to a stronger
form of dynamical localization than multiscale analysis (see
\cite{Aizenman, ASFH}) and has provided much shorter and more
transparent proofs in the lattice case, for example \cite{Graf}.

It was recently shown in \cite{AENSS} that all the main features
of the fractional moment approach also apply to continuum random
Schr\"odinger operators. This extension required substantial new
input from operator theory and harmonic analysis. The paper
\cite{AENSS} provides a framework of necessary and sufficient
criteria for localization in terms of fractional moment bounds,
which can be verified for a rather broad range of regimes.

One of our goals here is to complement the general framework from
\cite{AENSS} by focusing exclusively on presenting a short and
self-contained proof of localization properties via fractional
moments for one specific regime, where the technical effort
remains minimal.

For this we pick a fairly general setting we label the
\textit{fluctuation boundary regime}. This is described by a
random Schr\"odinger operator of Ander\-son-type in
$L^2(\D{R}^d)$, where our approach allows for quite arbitrary
background potentials and geometries of the random impurities,
provided the ground state energy is induced by rare events
(fluctuations) and therefore sensitive to changes in the random
parameters. The goal is to prove localization in the vicinity of
the bottom of the spectrum. Of course, various versions of the
fluctuation boundary regime have been studied in many works and we
borrowed the term from \cite{PF}.

Another motivation for our work is that we want to extend the
fractional moment method to situations where the random potential
does not satisfy a covering condition, i.e.\ where the individual
impurity potentials have small supports which do not cover all of
$\D{R}^d$. This condition, which was required for the technical
approach to the continuum found in \cite{AENSS}, is not natural in
the fluctuation boundary regime and should not be needed there as
has already been verified via multiscale analysis. Particularly
interesting examples are random surface potentials which act in a
small portion of space only. Nevertheless, they lead to a
fluctuation boundary by creating new ``surface spectrum'' below
the ``bulk spectrum''.

In our main result, Theorem~\ref{T1} below, the fluctuation
boundary regime will be described in form of an abstract
condition. For random surface potentials, which will be discussed
as an application, this condition follows in an appropriate
setting from a result proven in \cite{KW} in order to derive
Lifshits tails.
Another application concerns models with additional
random displacements as were originally studied in \cite{CH}.

Let us confess that we require absolutely continuous distribution
of random couplings. While it might be possible to relax this to
H\"older continuous distribution (as has been done in the lattice
case, e.g.\ \cite{ASFH}), the fractional moment method is so far less
flexible in that respect than the multiscale technique. In
particular, see the variant of multiscale analysis adapted to
Bernoulli-Anderson models recently developed in \cite{BK} and an
application of similar ideas to Poisson models announced in
\cite{GHK}.

%-------------------------------------------------------------%
\subsection{Results}

Let us now describe our results in more detail after introducing
some notation: On $\D{R}^d$ we often consider the supremum norm
$\abs{x}:=\max_{ i=1,\ldots,d}\abs{x_i}$ and write
\[
\Lambda_{r}(x):=\Bigl\lbrace y\in \D{R}^d: \abs{x-y} < \frac{r}{2}\Bigr\rbrace
\]
for the $d$-dimensional cube with sidelength $r$ centered at $x$.
For an open set $G\subset\D{R}^d$ we denote the restriction of the
Schr\"odinger operator $H$ to $L^2(G)$ with Dirichlet boundary
conditions by $H^G$. In our results we assume $d\leq 3$ and rely
upon the following assumptions, which guarantee self-adjointness
and lower semi-boundedness of all the Schr\"odinger operators
appearing in this paper:
\begin{enumerate}[{(A}1)]
\item
The background potential $V_0\in L^2_{\text{loc,unif}}(\D{R}^d)$
   is real-valued, $H_0:=-\Delta +V_0$.
\item
The set $\C{I}\subset\D{R}^d$, where the random impurities
   are located, is uniformly discrete, i.e., $\inf\{\abs{\alpha-\beta}:
   \alpha\neq\beta\in \C{I}\}=:r_\C{I}>0$.
\item
The random couplings $\eta_{\alpha}$, $\alpha\in\C{I}$,
   are independent random variables supported in $[0,\eta_{\max}]$ for some
   $\eta_{\max}>0$ and with absolutely continuous distribution of
   bounded density $\rho_\alpha$ with a uniform bound $\sup_\alpha \|\rho_\alpha
\|_\infty =: M_\rho<\infty$.\\
The single site potentials $U_{\alpha}$, $\alpha\in\C{I}$ satisfy
\[
          c_U\chi_{\Lambda_{r_U}({\alpha})}\le
          U_{\alpha}\leq C_U \chi_{\Lambda_{R_U}({\alpha})}
\]
 for  all ${\alpha}$ with  $c_U, C_U, r_U, R_U>0$ independent of $\alpha$.
\[
V_\omega(x)=\sum_{\alpha\in \C{I}} \eta_{\alpha }
   (\omega)U_{\alpha}(x)
\]
and
\[
H:=H(\omega):= H_0+V_\omega\text{ in } L^2(\D{R}^d).
\]
\end{enumerate}
The most important condition expresses the fact that the ground state
energy comes from those realizations of the potential that vanish on
large sets:
\begin{enumerate}[{(A}1)]
\addtocounter{enumi}{3}
\item
Denote $E_0:=\inf\sigma(H_0)\le\inf\sigma(H(\omega))$ and
   let
\[
H_F:=H_0+\eta_{\max}\sum_{\alpha\in \C{I}}U_{\alpha},
\]
 the subscript $F$ standing for full coupling.\\
Assume that $E_0$ is a
\textit{fluctuation boundary} in the sense that
\begin{enumerate}[(i)]
\item
$E_F:=\inf\sigma(H_F)>E_0$, and
\item[(ii)]
There is $m\in(0,2)$ and $L^*$ such
that for $m_d:=42\cdot d$, all $L\geq L^*$ and $ x\in \D{Z}^d$
\[
\D{P}\bigl(
     \sigma(H^{\Lambda_L(x)}(\omega)\cap[E_0,E_0+L^{-m}]
     \not=\emptyset\bigr)\leq L^{-m_d} .
\]
\end{enumerate}
\end{enumerate}

By $\chi_x$ we denote the characteristic function of the unit cube
centered at $x$. In the following it is understood that
$\chi_x(H^G-E-\ii\varepsilon)^{-1} \chi_y =0$ if $\Lambda_1(x) \cap
G$ or $\Lambda_1(y) \cap G$ have measure zero.

Our main result is

%------------------------%
\begin{thm} \label{T1}
   Let $d\leq 3$ and assume \emph{(A1)-(A4)}.  Then there exist
$\delta>0$, $0<s<1$,
   $\mu>0$ and $C<\infty$ such that for $I:=\croch{E_0,E_0+\delta}$, all
   open sets $G \subset\D{R}^d$ and $x,y\in \D{R}^d$,
\begin{equation}   \label{eq3}
\sup_{E\in I,\,\varepsilon >0}\expect (\norm{\chi_{x}
(H^G-E-\ii\varepsilon)^{-1}\chi_{y}}^{s}) \leq C\,\ee^{-\mu
\abs{x-y}}.
\end{equation}
\end{thm}
%------------------------%

Exponential decay of fractional moments of the resolvent as
described by \eqref{eq3} implies spectral and dynamical
localization in the following sense:

%------------------------%
\begin{thm}                             \label{T2}
Let $d\leq 3$, assume \emph{(A1)-(A4)} and let $I$ be given as in
Theorem~\ref{T1}. Then:
\begin{enumerate}[\rm(a)]
\item For all open sets $G\subset \D{R}^d$ the spectrum of $H^G$
in $I$ is almost surely pure point
with exponentially decaying eigenfunctions.
\item There are $\mu
>0$ and $C<\infty$ such that for all
   $x,y\in\D{R}^d$ and open $G\subset \D{R}^d$,
\begin{equation} \label{eq:dynloc}
\expect\bigl(\sup\norm{\chi_{x} g(H^G) P_I(H^G) \chi_y} \bigr) \leq
C\ee^{-\mu\abs{x-y}}.
\end{equation}
\end{enumerate}
where the supremum is taken over all Borel measurable functions $g$
which satisfy $|g|\leq 1$ pointwise and $P_I(H^G)$ is the spectral
projection for $H^G$ onto $I$.
\end{thm}
%------------------------%

Dynamical localization should be considered as the special case
$g(\lambda)=\ee^{\ii t\lambda}$ in (b), with the supremum taken over
$t\in \D{R}$.

The proof of Theorem~\ref{T1} is given in
Section~\ref{sec:mainproofs}. This will be done by a
self-contained presentation of a new version of the continuum
fractional moment method. While we use many of the same ideas as
\cite{AENSS}, due to the lack of a covering condition we can not
rely any more on the concept of ``averaging over local
environments", heavily exploited in \cite{AENSS}. It is
interesting to note that, in some sense, we instead use a global
averaging procedure. Technically, this actually leads to some
simplifications compared to the method in \cite{AENSS}, as
repeated commutator arguments can be replaced by simpler iterated
resolvent identities. We also mention that exponential decay in
\eqref{eq3} will follow from smallness of the fractional moments
at a suitable initial length scale (the localization length) via
an abstract contraction property.

As technical tools we need Combes-Thomas bounds (in operator norm
as well as in Hilbert-Schmidt norm) and a weak-$L^1$-type bound
for the boundary values of resolvents of maximally dissipative
operators, which is based on results from \cite{Naboko} and was
also central to the argument in \cite{AENSS}. We collect these
tools in an appendix.

That Theorem~\ref{T2} follows from Theorem~\ref{T1} was
essentially shown in \cite{AENSS}, Section~2. In
Section~\ref{sec:proofT2} below we will briefly discuss the
changes which arise due to our somewhat different set-up. In
particular, the argument in \cite{AENSS} for proving
\eqref{eq:dynloc} uses the covering condition
\begin{equation}
\label{eq:cover} 0< C_1 \leq\sum U_{\alpha} \leq C_2 < \infty
\end{equation}
in one occasion. But this is easily circumvented.

In Sections~\ref{sec:surface} and \ref{sec:displacement} we apply
our main result to concrete models by verifying assumption (A4)
for these models. In Section~\ref{sec:surface} we consider
Anderson-type random potentials supported in the vicinity of a
lower-dimensional surface. The ``usual'' fully stationary Anderson
model is considered in Section~\ref{sec:displacement}. The fact
that we don't have to assume a covering condition leads to high
flexibility in the geometry of the random scatterers. We could use
this to go for far reaching generalizations of Anderson models.
Instead, we restrict ourselves to the treatment of additional
random displacements as was done in \cite{CH}.

%-------------------------------------------------------------%
\subsection{Remarks}

We could have extended Theorem~\ref{T1} in at least two different
ways, but refrained from doing so to keep the proofs as
transparent as possible:
\begin{enumerate}[(i)]
\item
The restriction to $d\leq 3$ is not necessary. We use it
because in this case the abstract fractional moment bound in
Corollary~\ref{cor:fracmoments} is more directly applicable to our
proof of Theorem~\ref{T1} than in higher dimension (which
technically can be traced back to the fact that $\chi_{x}(-\Delta
+1)^{-1}$ is a Hilbert-Schmidt operator only for $d\leq 3$). In
higher dimension more iterations of resolvent identities would be
needed to yield the Hilbert-Schmidt multipliers required by
Corollary~\ref{cor:fracmoments}, leading to more involved
summations in the arguments of Section~\ref{sec:mainproofs}.
\item
Instead of bounded $U_\alpha$ we can work with relatively
   $\Delta$-bounded $U_\alpha$, i.e.\ allow for suitable $L^p$-type
   singularities in the single site potentials. In the course of our
   proofs they could be
   ``absorbed" into resolvents using standard arguments from
   relative perturbation theory.
\end{enumerate}

In principle, our arguments can also be used to prove localization
at fluctuation type band edges more general than the bottom of the
spectrum without using a covering condition as in \cite{AENSS}.
But this would require to be much more specific with settings and
assumptions and, in particular, with the geometry of the impurity
set. Inconvenience would also arise from having to work with
boundary conditions other than Dirichlet.

We mention that the applications in Section~\ref{sec:surface}
improve the results on continuum random surface potentials of
\cite{BdMS,KW}, obtained through the use of multiscale analysis:
\begin{enumerate}[(i)]
\item
The exponentially decaying correlations of the time evolution,
   shown as a special case of Theorem \ref{T2}(b), are stronger than
the dynamical bounds
   which follow from multiscale analysis.
\item
Due to the use of the recent result of \cite{KW} on Lifshitz
   tails for surface potentials, we do not need a condition on the
   smallness of the distribution of the $\eta_{\alpha}$ near the
   minimum of their support as in \cite{BdMS}, a progress that had been
   achieved in \cite{KW}.
\item
We can allow for more flexibility concerning the geometry of the
   scatterers.
\end{enumerate}

Of course, due to using fractional moments we cannot include
single site measures as singular as the ones considered in
\cite{BdMS,KW} but instead have to assume absolute continuity of
the $\eta_\alpha$.

%-------------------------------------------------------------%
\section{Localization near fluctuation boundaries}
\label{sec:mainproofs}

This section is entirely devoted to the proof
of Theorem 1. For a convenient normalization write
\begin{align*}
&\xi_{\alpha} (\omega) : = \eta_{\max} - \eta_{\alpha} (\omega)\\
&\text{for }
\omega = (\omega_{\alpha})_{\alpha \in \C{I}} =
(\eta_{\alpha}(\omega))_{\alpha\in \C{I}} \in \Omega:= [0,
\eta_{\max}]^{\C{I}},
\end{align*}
and denote the product measure $\otimes_{\alpha\in \C{I}}
\,\dd\eta_{\alpha}\rho_{\alpha}(\eta_{\alpha})$ on $\Omega$ by
$\D{P}$. We write
\[ W (x) : = W_{\omega} (x) : = \sum_{\alpha \in\C{I}}
\xi_{\alpha} (\omega) U_{\alpha} (x).
\]
Note that $W_{\omega} \geq 0$ and that
\[
H = H(\omega)= H_{F} - W_{\omega}.
\]
Fixing an open set $G \subset \D{R}^d$ we write
\begin{align*}
&R^G = R^G_z = (H^G - z)^{-1},\\
&R^G_F = R^G_{F,z} = (H^G_F-z)^{-1}
\end{align*}
whenever $z = E +\ii\varepsilon$. Since $H^G_F \geq H_F$ due to our
choice of Dirichlet boundary conditions, and $E_F = \inf \sigma
(H_F)$ we know that $(- \infty, E_F) \subset \rho (H^G_F)$.

The resolvent equation yields
\begin{equation} \label{eq:reseq}
R^G = R^G_F + R^G_F W R^G_F + R^G_F W R^G W R^G_F,
\end{equation}
an identity that will be used over and again. The other workhorse
result is the following averaging estimate, that follows from
Corollary \ref{cor:fracmoments} in the appendix below, taking into
account the uniform boundedness of the densities $\rho_\alpha$.

%------------------------%
\begin{lem}\label{workhorse}
For all $s\in [0,1)$ there is $c(s)$ such that
\begin{align*}
\int\dd\eta_{\alpha} \rho_{\alpha} (\eta_{\alpha}) \int\dd
\eta_{\beta} \rho_{\beta} (\eta_{\beta})
& \|M_1 U_{\alpha}^{1/2}
(H^G-E-\ii\varepsilon)^{-1} U_{\beta}^{1/2} M_2 \|^s_{\HS}\notag\\
& \leq c(s) \|M_1\|^s_{\HS}\|M_2\|^s_{\HS}.
\end{align*}
\end{lem}
%------------------------%

As a warm-up, we prove boundedness of fractional moments:

%------------------------%
\begin{lem}\label{L3}
Let $E_1 < E_F$, $I = [E_0, E_1]$ and $s \in [0,1)$.
Then
\begin{equation} \label{eq:boundedfm}
\sup\{ \expect\| \chi_x R^G_{E + i \varepsilon} \chi_y \|^s\mid E \in
I, \varepsilon > 0, x,y \in \D{R}^d, G\subset \D{R}^d\,
\text{open}\} < \infty\:.
\end{equation}
\end{lem}
%------------------------%

%------------------------%
\begin{proof}
We use \eqref{eq:reseq} above and write, suppressing the superscript $G$ and
the subscript $z=E+\ii\varepsilon$ mostly:
\[
  \chi_x R \chi_y = \chi_x R_F \chi_y + \chi_x R_F W R_F \chi_y
+ \chi_x R_F W R W R_F \chi_y.
\]
The first two terms on the r.h.s.\ of this equation obey an
exponential bound due to the Combes-Thomas estimate, see
subsection \ref{CT} below:
\[
\| \chi_x R_F \chi_y \| \leq c\,\ee^{- \mu_0|x-y|}
\]
and
\begin{align*}
\|\chi_x R_F W R_F \chi_y \|
& \leq\eta_{\max} \sum_{\alpha\in
\C{I}} \|\chi_x R_F U_{\alpha}^{1/2}\|\cdot\|U_{\alpha}^{1/2} R_F\chi_y \| \notag\\
&\leq
C \sum_{\alpha\in \C{I}}
\ee^{-\mu_0|x-\alpha|} \ee^{-\mu_0 |\alpha-y|} \leq C \ee^{-\mu_1 |x-y|}
\end{align*}
with $\mu_0$ and $\mu_1 = \mu_0/2$ depending on $E_1$ only. In the
last estimate we have used that $\C{I}$ is uniformly discrete.

For the third term, expand $W = \sum_{\alpha} \xi_{\alpha}
U_{\alpha}$ and use the boundedness of the $\xi_{\alpha}$ and the
fact that
\[
\Bigl(\sum a_n\Bigr)^s \leq\sum a^s_n
\]
to estimate
\[
  \| \chi_x R_F W R W R_F \chi_y \|^s \leq c \sum_{\alpha,\,\beta \in \C{I}}
\| \chi_x R_F U_{\alpha} R U_{\beta} R_F \chi_y \|^s.
\]
We now fix $\alpha, \beta \in \C{I}$ and use the workhorse
Lemma~\ref{workhorse} to conclude
\begin{align*}
\int\dd\eta_{\alpha} \rho_{\alpha} (\eta_{\alpha})  \int\dd\eta_{\beta}
\rho_{\beta} (\eta_{\beta})
& \| \chi_xR_F U_{\alpha} R U_{\beta}
R_F \chi_y \|^s \notag \\
& \leq
c(s) \| \chi_x R_F U_{\alpha}^{1/2}\|^s_{\HS} \|
U_\beta^{1/2} R_F \chi_y \|^s_{\HS} \notag \\
& \leq
c(s) \cdot \ee^{-s \mu_0 |x-\alpha|} \cdot \ee^{-s \mu_0
|y-\beta|}
\end{align*}
by the HS-norm Combes-Thomas bound from
Proposition~\ref{prop:CTHS} and since $\dist(x,\supp U_{\alpha})
\geq |x-\alpha| -R_U $ where $R_U$ majorizes the size of the
support of $U_{\alpha}$ according to assumption (A3).

Note that here and in the following we use the convention that $c$,
$c(s)$, etc.\ denote constants that only depend on non-crucial
quantities and may change from line to line. In particular, the
constants are independent of $\varepsilon > 0$ and the random
background.

Now, we can sum up the last terms and get the assertion.
\end{proof}
%------------------------%

%------------------------%
\begin{rems*}
(i)
In this proof it is still quite easy to see how to extend
to arbitrary dimension through iterations of the resolvent
identity. It will be harder to keep track of this later.

(ii)
Note that due to the $\alpha, \beta$-summations, averaging
over the $\eta_{\alpha}$ is required for all $\alpha$, i.e., is
global. In \cite{AENSS}, due to the covering condition, an
argument is provided that only requires averaging over local
environments of $x$ and $y$ and proves Lemma~\ref{L3} for
arbitrary finite intervals $I = [E_0, E_1]$, i.e.\ without
requiring $E_1 < E_F$.

(iii) The above proof shows that \eqref{eq:boundedfm} also holds
in HS-norm, but this will not be used below.
\end{rems*}
%------------------------%

We will now start an iterative procedure that will show
exponential decay of $\expect(\|\chi_x R \chi_y \|^s)$ in $|x-y|$ for
energies sufficiently close to $E_0$. Clearly, it suffices to
consider $x,y \in \D{Z}^d$. In view of the preceding lemma the
following quantity is finite:
\[
\tau_{x,y} : = \sup\{ \expect\| \chi_x R^G_{E + i \varepsilon} \chi_y \|^s
\mid E \in I,\,\varepsilon > 0\text{  and  }G\subset \D{R}^d\,
\text{open}\}\,.
\]
Moreover, we should actually keep in mind the dependence on the
interval $I=[E_0, E_1]$. In fact, $E_1$ will later be chosen small
enough.

In order to use that $E_0$ appears rarely as an eigenvalue for
boxes of side length $L$ we exploit the resolvent identity and
what is sometimes called the Simon-Lieb inequality in a way
visualized in Figure 1!
\begin{center}
\begin{figure}[h]
   \centering
         \includegraphics[width=12cm]{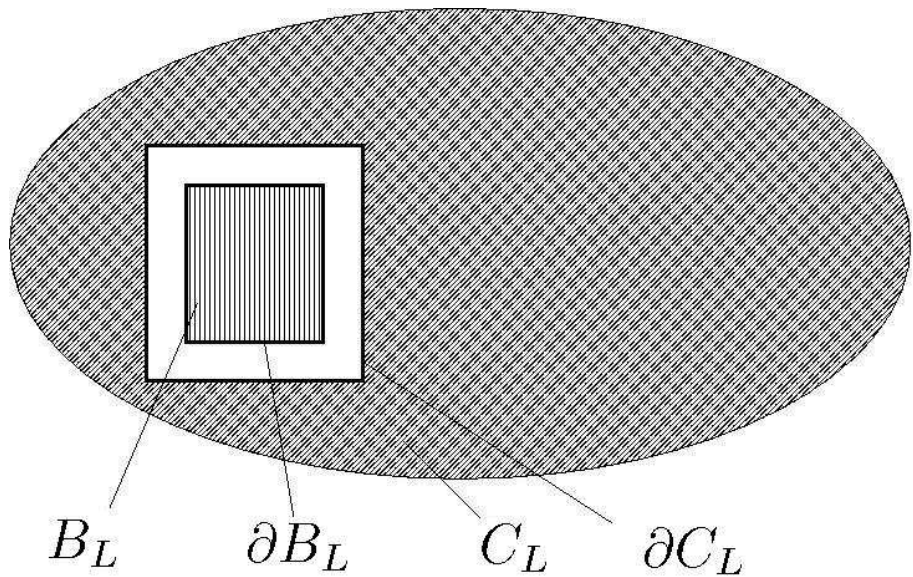}
         \caption{The geometry of the induction step}
\end{figure}
\end{center}
Consider
\begin{align*}
&B_L := \Lambda_L (x) \cap G,\\
&\partial B_L = (\Lambda_L (x) \setminus
\Lambda_{L-2} (x)) \cap G,\text{ and } \chi_L^- := \chi_{\partial B_L}.
\end{align*}
Furthermore, with $R_U$ as in assumption (A3), define
\begin{align*}
&
C_L :=  G\setminus\overline{\Lambda_{2R_U+L}(x)},\\
&
\partial C_L = \bigl(\Lambda_{2R_U+L+2} (x)
\setminus\overline{\Lambda_{2R_U+L} (x)}\,\bigr) \cap G,\text{ and } \chi_L^+ :=
\chi_{\partial C_L}.
\end{align*}
The geometry is chosen in such a way that $R^{B_L}$ and $R^{C_L}$
are stochastically independent. For $R^{B_L}$ we can use the
fluctuation boundary assumption to get small fractional moments
and the right size of $L$ will be adjusted.
But all that later...

Thus, by the Simon-Lieb inequality (e.g.\ \cite{Stollmann},
Sect.~2.5)
\[
\| \chi_x R^G \chi_y \| \leq C \| \chi_x R^{B_L} \chi_L^- \| \cdot
\| \chi_L^{-} R^G \chi_{L}^{+} \| \cdot \| \chi_{L}^+ R^{C_L}
\chi_y \| \tag{SLI}
\]
where $C$ only depends on
$\sup_{\{\eta_{\alpha}\mid\alpha\in\C{I}\}}\norm{V}_{\infty}$ and the
interval $I$.

The basic idea for proving exponential decay of $\tau_{x,y}$ is to
establish a recurrence inequality for energies sufficiently close
to $E_0$. This recurrence  inequality is described in Proposition
\ref{master} below and allows to apply a discrete Gronwall-type
argument found in Lemma \ref{lem.estimate.operator.A} below. To
this end we exploit smallness of fractional moments of the first
factor on the r.h.s.\ of (SLI) for energies close to $E_0$ and
sufficiently large, but fixed, $L$: This will follow from (A4)(ii)
as is presented in the following Lemma \ref{startup}. Fractional
moments of the second factor are bounded due to Lemma~\ref{L3} (up
to a polynomial factor in $L$). Finally, we use the third factor to
start an iteration (with $x$ replaced by sites $x'$ covering the
layer $\Lambda_{L+R_U+2}\setminus\Lambda_{L+R_U}$). By
construction, the first and third factor on the r.h.s.\ of (SLI)
are probabilistically independent. Unfortunately, the second
factor introduces a correlation which prevents us from simply
factoring the expectation. We will rely on a version of the
re-sampling procedure developed in \cite{AENSS} to solve this
problem. Moreover, we will not use Lemma~\ref{L3}, but apply
Lemma~\ref{workhorse} directly to bound certain conditional
expectations. This will result in Proposition~\ref{master} below.

%------------------------%
\begin{lem}\label{startup}
For $m$ as in \emph{(A4)} and $s\in (0,\frac13)$ there is
$L^*=L^*(m,s)$ such that for all $L\geq L^*$, open $B\subset
\Lambda_L(x)$, $E\in I:=[E_0,E_0+ L^{-\frac12 m}]$, $\varepsilon
>0$ and $u,v\in \D{Z}^d$ with $| u-v|\ge\frac{L}{4}$ we have
\[
\expect(\| \chi_u(H^B-E-\ii\varepsilon)^{-1}\chi_v\|^s)\le
L^{-\frac{1}{2}m_d} ,
\]
where $m_d=42\cdot d$.
\end{lem}
%------------------------%

%------------------------%
\begin{proof}
Divide $\Omega$ into the good and bad sets
\[
\Omega_{\text{good}}:= \{ \omega\mid\dist( \sigma (H^B), E_0) >
  L^{-m} \},\quad\Omega_{\text{bad}} =\Omega\setminus \Omega_{\text{good}} .
\]
Since $H^B\geq H^{\Lambda_L(x)}$ by our choice of Dirichlet
boundary conditions, (A4) implies that
\[
\D{P} (\Omega_{\text{bad}}) \leq L^{- m_d} .
\]
We split the expectation into contributions from the good and bad
sets: By the improved Combes-Thomas bound Subsection \ref{CT} we
get, for $\omega \in \Omega_{\text{good}}$, $E \in I$:
\[
\| \chi_u R^B_{E+\ii\varepsilon} \chi_v \|^s \leq C L^{\frac12 m s}
\ee^{-c s |u-v| L^{-\frac12 m}}\:.
\]
This
gives a uniform bound of the same type for the expectation over $
\Omega_{\text{good}}$. For the bad set H\"{o}lder with $t\in (s,1)$ gives
\begin{align*}
\expect( \| \chi_u R^B_{E+\ii\varepsilon} \chi_v \|^s
\chi_{\Omega_{\text{bad}}})
&\leq
\left(\expect( \| \chi_u
R^B_{E+\ii\varepsilon} \chi_v \|^t)\right)^
{\frac{s}{t}} \D{P} (\Omega_{\text{bad}})^{1-\frac{s}{t}}\\
&\leq
c(t)^{\frac{s}{t}} L^{(1-\frac{s}{t})m_d}.
\end{align*}
Now we choose $t=\frac12(s+1)$ so that $(1-\frac{s}{t})>\frac12$
if $s<\frac13$. Putting things together, we get
\[
\expect( \| \chi_u R^B_{E+\ii\varepsilon} \chi_v \|^s ) \le
C(s)\Bigl( L^{\frac12 m s} \ee^{-c s |u-v| \cdot L^{-\frac12 m}} +
L^{(1-\frac{s}{t})m_d}\Bigr) .
\]
If $L$ is large enough we can use $\frac12 m<1$ and
$|u-v|\geq\frac{L}{4}$ to see that the r.h.s.\ is bounded as asserted.
\end{proof}
%------------------------%

The exponential decay of the $\tau_{x,y}$ will follow from the
following result, whose proof will take most of the present
section.

%------------------------%
\begin{prop}\label{master}
There exist $L^*$, $\kappa>0$, $c>0$ and $C>0$, all depending on
$s,m,R_U,r_U,M_\rho,E_0,E_F,\eta_{\max}$, such that for $L\geq L^*$
and $I=[E_0,E_0+L^{-\frac12 m}]$ the above defined $\tau_{x,y}$
satisfy:
\begin{equation} \label{eq:recursion}
\tau_{x,y}\leq L^{-2d-\kappa} \sum_{x',\,y'\in \D{Z}^d}
\ee^{-c(|x-x'|+|y-y'|)/L}\,\tau_{x',y'} + C \ee^{-c|x-y|/L} .
\end{equation}
\end{prop}
%------------------------%

%------------------------%
\begin{proof}[Proof of Proposition~\ref{master}]
We now restrict to the energy interval
$I = [E_0, E_0 +
L^{-\frac12 m}]$ assuming $L$ is large enough to guarantee that
$I\subset [E_0,E_F)$. Using (SLI) above and denoting
\begin{align*}
T_{x,L}
&= \chi_x R^{B_L} \chi_L^{-} , \\
S_{x,L}
&= \chi_L^{-}R^G\chi_L^{+}  ,\\
Q_{x,L}
&= \chi_L^+ R^{C_L} \chi_y
\end{align*}
we get that
\[
\expect(\| \chi_x R^G \chi_y \|^s) \leq C\expect(\| T_{x,L} \|^s \|
S_{x,L} \|^s \| Q_{x,L} \|^s).
\]
Note that $\| T_{x,L} \|^s$ and $\| Q_{x,L} \|^s$ are
stochastically independent. Unfortunately, they are correlated via
$\| S_{x,L} \|^s$.

Fix $s \in (0, \frac{1}{3})$ to estimate $\expect(\| T_{x,L} \|^s)$.
Using the preceding Lemma, we get that
\begin{align*}
\expect(\| T_{x,L} \|^s)
&\leq\sum_{z \in \supp \chi_L^-}
\expect(\| \chi_x R^{B_{L}} \chi_z \|^s)\\
&\leq C L^{d-1} \cdot L^{-\frac{1}{2} m_d} ,
\end{align*}
for $L$ large enough. We get that
\[
\expect(\| T_{x,L} \|^s) \leq L^{d-\frac12 m_d} .
\]
We can now expand $S_{x,L}$ to split off a uniformly bounded (in
$\omega$) term:
\[
S_{x,L} = \underbrace{\chi_{L}^{-} R^G_F \chi_{L}^{+} +
\chi_L^- R^G_F W R^G_F \chi_L^{+}}_{S_{1,L}}
+ \underbrace{\chi_L^{-} R^G_F W R^G W R^G_F \chi_L^{+}}_{S_{2,L}}.
\]
Since $I \subset [E_0, E_F)$ we have
that $\| S_{1,L} \|^s$ is uniformly bounded. Thus, we get
\begin{align*}
\expect(\| \chi_x R^G_{E+\ii\varepsilon} \chi_y \|^s)
&\leq
C\bigl(\expect(\| T_{x,L} \|^s \cdot \| Q_{x,L} \|^s) +  \Sigma_2\bigr) \\
&=
C\bigl( \expect( \| T_{x,L} \|^s) \cdot\expect(\| Q_{x,L} \|^s) +
\Sigma_2\bigr)
\end{align*}
as $\| T_{x,L} \|^s$ and $\| Q_{x,L} \|^s$ are independent. Here
\[
\Sigma_2 := \expect( \| T_{x,L} \|^s \| S_{2,L} \|^s Q_{x,L}
\|^s).
\]
Expanding $\chi_L^{+}$ we get, for some $c>0$, that
\[
\ldots\leq C L^{d-\frac12 m_d} \sum_{x' \in \partial C_L} \ee^{- c
|x-x'|/L} \expect(\| \chi_{x'}  R^{C_{L}} \chi_y \|^s) + C\,
\Sigma_2,
\]
whence
\begin{equation} \label{eq:star}
  \tau_{x,y} \leq L^{d-\frac12 m_d} \sum_{x' \in\partial C_L}
\ee^{- c |x-x'|/L}\,\tau_{x',y} + C \sup_{\substack{E\in I,\,\varepsilon>0\\
                                                                                   G \subset \D{R}^d}} \Sigma_2 .
\end{equation}
To estimate $\Sigma_2$ we begin by expanding
\begin{align} \label{eq.first.factor.expansion}
T_{x,L}
&= \chi_x R^{B_L}\chi_L^{-}\notag\\
&= \chi_x R^{B_L}_F\chi_L^{-}+
\chi_x R^{B_L}_F WR^{B_L}_F\chi_L^{-}
+ \chi_x R^{B_L}_F WR^{B_L}WR^{B_L}_F\chi_L^{-}.
\end{align}
Since $I$ has positive distance from $\sigma(H_F)$, we have the
Combes-Thomas bound $C\ee^{-\mu_0L/2}$ for the norm of the first two
terms on the r.h.s.\ of \eqref{eq.first.factor.expansion}, see
Appendix~\ref{CT}. Here $C<\infty$ and $\mu_0>0$ are uniform in
the randomness, $E\in I$, $\varepsilon>0$ and $x\in\D{Z}^d$.
Expanding the third term and using boundedness of the $\xi$'s
yields
\[
\| T_{x,L}\|^s \leq C\biggl( \ee^{-\mu_0\cdot s\cdot\frac{L}{2}} +
\sum_{\beta,\gamma\in \C{I} \cap \Lambda_{L+R_U}(x)} \|
T_{\beta,\gamma}\|^s\biggr),
\]
setting $T_{\beta,\gamma} = \chi_x R^{B_L}_F U_{\beta}R^{B_L}
U_{\gamma}R^{B_L}_F\chi_L^{-}$, and only summing over those
$\beta, \gamma$ for which the corresponding $U$-terms touch $B_L$.

A similar argument applied to $Q_{x,L}$ leads to
\[
\| Q_{x,L} \|^s \leq C\biggl( \ee^{-\mu_0\cdot s\cdot (|x-y|
-\frac{L}{2})}+\sum_{\beta',\gamma'\in \C{I} \cap
\Lambda_{L+R_U}^c(x)} \| Q_{\beta',\gamma'} \|^s\biggr),
\]
where we have chosen $Q_{\beta',\gamma'} = \chi_L^{+}R^{C_L}_F
U_{\gamma'}R^{C_L}U_{\beta'} R^{C_L}_F\chi_y$.

Finally, expand
\[
S_{2,L} = \chi_L^{-} R^G_F W R^G W R^G_F \chi_L^+
=\sum_{\alpha,\alpha'\in \C{I}} S_{\alpha,\alpha'},
\]
where $S_{\alpha,\alpha'} = \chi_L^{-} R^G_F\xi_\alpha U_\alpha
R^G\xi_{\alpha'} U_{\alpha'} R^G_F \chi_L^+$.

Combining all this we get that
\begin{align*}
\Sigma_2
&\leq
C\biggl( \ee^{-\mu_0\cdot s\cdot\frac{L}{2}}
\sum_{\alpha,\alpha'}\expect\| S_{\alpha,\alpha'} \|^s
\ee^{-\mu_0\cdot s\cdot (|x-y| -\frac{L}{2})}\\
&\qquad\quad
+ \ee^{-\mu_0\cdot
s\cdot\frac{L}{2}}\sum_{\alpha,\alpha',\beta',\gamma'} \expect(\|
S_{\alpha,\alpha'} \|^s\cdot \| Q_{\beta',\gamma'}\|^s)\\
&\qquad\quad
+ \sum_{\alpha,\alpha',\beta,\gamma}\expect(\|
T_{\beta,\gamma}\|^s\cdot \| S_{\alpha,\alpha'} \|^s)
\ee^{-\mu_0\cdot s\cdot (|x-y| -\frac{L}{2})} \\
&\qquad\quad
+ \sum_{\alpha,\alpha',\beta,\beta',\gamma,\gamma'}
\expect(\| T_{\beta,\gamma}\|^s\cdot \| S_{\alpha,\alpha'}
\|^s\cdot  \| Q_{\beta',\gamma'}\|^s)\biggr) .
\end{align*}
The most complicated of these terms is the last one; it will be
obvious how to estimate the first three once we have established a
bound for the last one according to the assertion of
Proposition~\ref{master}. Thus we have to estimate
\[
\Sigma_3:= \sum_{\alpha,\alpha',\beta,\beta',\gamma,\gamma'}
A_{\alpha,\alpha',\beta,\beta',\gamma,\gamma'},
\]
where
\[
A_{\alpha,\alpha',\beta,\beta',\gamma,\gamma'} = \expect\left(\|
T_{\beta,\gamma}\|^s\cdot \| S_{\alpha,\alpha'} \|^s\cdot \|
Q_{\beta',\gamma'}\|^s \right) .
\]
If it weren't for the $S_{\alpha,\alpha'}$-terms, the
$T_{\beta,\gamma}$ and $Q_{\beta',\gamma'}$ would be independent,
leading to an estimate like in \eqref{eq:star} above. We will
reinforce a certain kind of independence through
\textit{re-sampling}. For fixed
\[
\C{J}:=\{\alpha,\alpha',
\gamma,\gamma'\}
\]
we introduce new independent random variables
$\widehat{\xi}_j$, $j\in \C{J}$, independent of the $\xi_{\zeta}$,
$\zeta\in \C{I}$, and with the same distribution as the
$\xi_{\zeta}$. We denote the corresponding space by
$\widehat{\Omega}$, the corresponding probability by
$\widehat{\D{P}}$ and the expectation with respect to
$\widehat{\D{P}}$ by $\expecthat$. Consider
\[
\widehat{H}(\omega,\widehat{\omega})= H(\omega)+ \underbrace{
\sum_{j\in \C{J}}(\xi_j(\omega)
-\widehat{\xi}_j(\widehat{\omega})) U_j}_{\widehat{W}}
\]
and note that $\widehat{H}$ doesn't depend on the $\xi_j$, $j\in
\C{J}$. The resolvent identity for
$\widehat{R}^G_z=(\widehat{H}^G-z)^{-1}$ gives
\[
R^G_z=\widehat{R}^G_z+\widehat{R}^G_z \widehat{W}R^G_z .
\]
We insert this for $T_{\beta,\gamma}$ and $Q_{\beta',\gamma'}$ and
get
\[
T_{\beta,\gamma}= \underbrace{\chi_x  R^{B_L}_F
U_\beta\widehat{R}^{B_L}U_ \gamma
R^{B_L}_F\chi_L^{-}}_{\widehat{T}_{\beta,\gamma}}
+\underbrace{\chi_xR^{B_L}_F
U_\beta\widehat{R}^{B_L}\widehat{W}R^{B_L}U_ \gamma
R^{B_L}_F\chi_L^{-}}_{\widetilde{T}_{\beta,\gamma}}
\]
and, similarly,
\[
Q_{\beta',\gamma'}=\widehat{Q}_{\beta',\gamma'}+
\widetilde{Q}_{\beta',\gamma'} .
\]
Now we can estimate
\begin{equation} \label{eq:starstar}
A_{\alpha,\alpha',\beta,\beta', \gamma,\gamma'}\le
\expecthat\expect\left[ (\|  \widehat{T}_{\beta,\gamma}\|^s+
\| \widetilde{T}_{\beta,\gamma} \|^s) \| S_{\alpha,\alpha'}
\|^s(\|\widehat{Q}_{\beta',\gamma'}\|^s+ \|
\widetilde{Q}_{\beta',\gamma'}\|^s)\right] .
\end{equation}
This gives a sum of four terms we have to control. Let's start
with the easiest one
\[
A^1_{\alpha,\alpha',\beta,\beta',
\gamma,\gamma'}:=\expecthat\expect\bigl[ \|
\widehat{T}_{\beta,\gamma}\|^s\| S_{\alpha,\alpha'} \|^s
\|\widehat{Q}_{\beta',\gamma'}\|^s\bigr] .
\]
Denote
\[
\expect(X|\alpha,\alpha')=\int\dd\xi_\alpha\rho_\alpha(\xi_\alpha)
\int\dd\xi_{\alpha'}\rho_{\alpha'}(\xi_{\alpha'})\,X(\xi)
\]
for a random variable on $\Omega\times\widehat{\Omega}$, so that
$\expect(X|\alpha,\alpha')$ is nothing but the conditional
expectation with respect to the $\sigma$-field generated by the
family $(\xi_\beta\mid\beta\in\C{I}\setminus\{\alpha,\alpha'\})$.
According to the usual rules for conditional expectations:
\begin{align} \label{eq:starstarstar}
A^1_{\alpha,\alpha',\beta,\beta', \gamma,\gamma'}
&=
\expecthat\expect\left[\expect( \|
\widehat{T}_{\beta,\gamma}\|^s\| S_{\alpha,\alpha'} \|^s
\|\widehat{Q}_{\beta',\gamma'}\|^s|\alpha,\alpha')\right] \notag \\
&=
\expecthat\expect\left[ \|
\widehat{T}_{\beta,\gamma}\|^s\|\widehat{Q}_{\beta',\gamma'}\|^s
\expect(\| S_{\alpha,\alpha'} \|^s|\alpha,\alpha')\right]
\end{align}
since the $\widehat{T}$ and $\widehat{Q}$ are independent of
$\xi_\alpha,\xi_{\alpha'}$. Using the workhorse
Lemma~\ref{workhorse} and the Combes-Thomas estimate
Proposition~\ref{prop:CTHS} we get
\begin{align*}
\expect(\| S_{\alpha,\alpha'} \|^s|\alpha,\alpha')
&\leq
c(s) \|
\chi_L^{-} R^G_F U_\alpha^{\frac12}\|^s_{\HS}
\| U_{\alpha'}^{\frac12} R^G_F \chi_L^{+} \|^s_{\HS}\\
&\leq
c(s) L^{2s(d-1)} \ee^{-\mu_1s(|\frac{L}{2}-|\alpha-x|| +
|\frac{L}{2}-|\alpha'-x||)},
\end{align*}
where the extra $L^{2s(d-1)}$ term comes from covering $\partial
B_L$ and $\partial C_L$. We have
\[
\expecthat\expect\left[  \|
\widehat{T}_{\beta,\gamma}\|^s\|\widehat{Q}_{\beta',\gamma'}\|^s\right]
= \expect\left[  \|
{T_{\beta,\gamma}}\|^s\|{Q_{\beta',\gamma'}}\|^s\right] = \expect
\left[  \|  {T_{\beta,\gamma}}\|^s\right]\expect\left[
\|{Q_{\beta',\gamma'}}\|^s\right]
\]
since the $\widehat{\xi}$'s have the same distribution as the
$\xi$'s and the $T$'s and $Q$'s are independent. Inserting into
\eqref{eq:starstarstar} gives
\[
A^1_{\alpha,\alpha',\beta,\beta', \gamma,\gamma'}
\leq
c(s) L^{2s(d-1)} \ee^{-\mu_1s(|\frac{L}{2}-|\alpha-x|| +
|\frac{L}{2}-|\alpha'-x||)}
\expect\left[  \|
{T_{\beta,\gamma}}\|^s\right]\expect\left[
\|{Q_{\beta',\gamma'}}\|^s\right].
\]
We will now treat the latter two terms separately:

%------------------------%
\subsubsection*{Step 1}
Denote by $Z(\gamma')=\{ y'\in\D{Z}^d\mid
\chi_{y'}\cdot U_{\gamma'}\not= 0\}$ those lattice points whose
1-cubes  support $U_{\gamma'}$. By Combes-Thomas once again:
\begin{align*}
\|Q_{\beta',\gamma'}\|^s
&=
\|
\chi_L^{+}R^{C_L}_FU_{\gamma'}R^{C_L}U_{\beta'}R^{C_L}_F\chi_{y}\|^s\\
&\leq
C\sum_{x'\in Z(\gamma')} \sum_{y'\in Z(\beta')} \|
\chi_{x'}R^{C_L}\chi_{y'} \|^s\ee^{-\mu_1s(|x-x'|-L)}
\ee^{-\mu_1s|y-y'|} .
\end{align*}
By assumption on the size of the support of $U_{\gamma'}$ we see
that $\# Z(\gamma')$ is uniformly bounded. This and uniform
discreteness of $\C{I}$ gives
\[
\sum_{\beta',\gamma'} \expect\|Q_{\beta',\gamma'}\|^s
\leq  C\sum_{x',\,y'\in \D{Z}^d \cap C_L}
\ee^{-\mu_1s(|x-x'|-\frac{L}{2})} \ee^{-\mu_1s|y-y'|}\,\tau_{x',y'} .
\]

%------------------------%
\subsubsection*{Step 2}
For the $T_{\beta,\gamma}$-term we have
\begin{align*}
\| T_{\beta,\gamma} \|^s
&=
\| \chi_xR^{B_L}_FU_\beta R^{B_L}U_\gamma R^{B_L}_F\chi_L^{-}\|^s \\
&\leq
C\sum_{u\in Z(\beta) \cap B_L} \sum_{v\in Z(\gamma) \cap
B_L} \| \chi_x R^{B_L}_F\chi_u\|^s \|\chi_u R^{B_L}\chi_v\|^s
\|\chi_v R^{B_L}_F\chi_L^{-}\|^s .
\end{align*}
If $|u-v|\geq\frac14 L$, Lemma~\ref{startup} gives
\[
\expect(\| \chi_u R^{B_L}\chi_v\|^s)
\leq
C\cdot L^{-\frac12 m_d} .
\]
If, on the other hand, $|u-v|\leq\frac14 L$ then $\dist(v,\partial
B_L)\geq\frac18 L$ or $|x-u|\geq\frac18 L$, so that the uniform
bound of Lemma~\ref{L3} for $\expect(\| \chi_u R^{B_L}\chi_v \|^s)$
together with the Combes-Thomas bound for $\| \chi_v
R^{B_L}_F\chi_L^{-}\|^s$, resp.\ $\| \chi_x R^{B_L}_F\chi_u\|^s$
gives, for $L$ large enough,
\begin{align*}
\expect\bigl(\| \chi_x R^{B_L}_F\chi_u\|^s \|\chi_u
R^{B_L}\chi_v\|^s
\|\chi_v R^{B_L}_F\chi_L^{-}\|^s \bigr)
&\leq
C  \ee^{-\frac18\mu_0s L}\\
&\leq
L^{-\frac12 m_d} .
\end{align*}
Combined we get that, again for $L$ sufficiently large,
\[
\sum_{\beta,\gamma}\expect(\| T_{\beta,\gamma}\|^s) \leq C
L^{2d-\frac12 m_d} ,
\]
where an extra factor $L^{2d}$ arises through the number of terms
considered.

Joining Step~1, Step~2 and the bound
\[
\sum_{\alpha, \alpha'} \ee^{-\mu_1s(|\frac{L}{2}-|\alpha-x|| +
|\frac{L}{2}-|\alpha'-x||)} \leq C(s) L^{2d}
\]
we arrive at
\begin{align*}
&\sum_ {\alpha,\alpha',\beta,\beta',
\gamma,\gamma'}A^1_{\alpha,\alpha',\beta,\beta', \gamma,\gamma'}\\
&\qquad\qquad\quad
\leq
C(s) L^{6d-\frac12 m_d} \sum_{x',\,y'\in\D{Z}^d \cap C_L} \ee^{-\mu_1
s(|x-x'|-\frac{L}{2})} \ee^{-\mu_1 s|y-y'|}\,\tau_{x',y'},
\end{align*}
which is a contribution to $\Sigma_3$ (and therefore $\Sigma_2$)
bounded by one of the type asserted in Proposition~\ref{master}.

A look back at \eqref{eq:starstar} shows that we still have to
estimate three terms similar to $A^1_{\alpha,\alpha',\beta,\beta',
\gamma,\gamma'}$ of which the last one,
\[
A^4_{\alpha,\alpha',\beta,\beta', \gamma,\gamma'}
:=\expecthat\expect\left[ \|
\widetilde{T}_{\beta,\gamma}\|^s\| S_{\alpha,\alpha'} \|^s
\|\widetilde{Q}_{\beta',\gamma'}\|^s\right]
\]
is the most complicated one. Using Steps 1 and 2 above as well as
the steps below it will be
clear how to treat the two remaining terms.

%------------------------%
\subsubsection*{Step 3}
We start taking the conditional expectation:
\begin{align*}
A^4_{\alpha,\alpha',\beta,\beta', \gamma,\gamma'}
&=\expecthat\expect\left[\expect(\|
\widetilde{T}_{\beta,\gamma}\|^s\| S_{\alpha,\alpha'} \|^s
\|\widetilde{Q}_{\beta',\gamma'}\|^s| \alpha,\alpha',
\gamma,\gamma')\right]\\
&\leq
\expecthat\expect\Bigl\lbrack\expect(\|
\widetilde{T}_{\beta,\gamma}\|^{3s} |\alpha,\alpha',
\gamma,\gamma')^\frac13\\
&\qquad\quad
\cdot\expect(\| S_{\alpha,\alpha'} \|^{3s} |\alpha,\alpha',
\gamma,\gamma')^\frac13\cdot\expect(\|\widetilde{Q}_{\beta',\gamma'}\|^{3s}
  |\alpha,\alpha',
\gamma,\gamma')^\frac13\Bigr\rbrack
\end{align*}
by H\"{o}lder's inequality. Like above, the middle term can, up to
$CL^{2s(d-1)}$, be estimated by
\[
e_{\alpha,\alpha'}:= \ee^{-\mu_1 s ||\alpha-x|-\frac{L}{2}|}
\ee^{-\mu_1 s ||\alpha'-x|-\frac{L}{2}|} .
\]
Recall that
\begin{align*}
\|\widetilde{Q}_{\beta',\gamma'}\|^{3s}
&=
\bigl\lVert
\chi_L^{+}{R}^{C_L}_FU_{\gamma'}{R}^{C_L} \sum_{j\in \C{J}
\setminus \{\gamma\}} (\xi_j-\widehat{\xi}_j) U_j\widehat{R}^{C_L}
U_{\beta'}R^{C_L}_F\chi_y\bigr\rVert^{3s}\\
&\leq
C\cdot \sum_{j\in \C{J}\setminus \{\gamma\}}\|
\chi_L^{+}{R}^{C_L}_FU_{\gamma'}{R}^{C_L} U_j\widehat{R}^{C_L}
U_{\beta'}R^{C_L}_F\chi_y\|^{3s},
\end{align*}
where $\gamma$ can be excluded from the summation as $U_{\gamma}$
doesn't touch $C_L$. Integration over $\xi_j$ and $\xi_{\gamma'}$
gives a uniform bound by the workhorse Lemma~\ref{workhorse}:
\begin{align*}
\expect( \|\widetilde{Q}_{\beta',\gamma'}\|^{3s}|
&\alpha,\alpha',
\gamma,\gamma')\\
&\leq\sum_{j\in \C{J} \setminus \{\gamma\}}\expect(\|
\chi_L^{+}{R}^{C_L}_FU_{\gamma'}{R}^{C_L} U_j\widehat{R}^{C_L}
U_{\beta'}R^{C_L}_F\chi_y\|^{3s} |\alpha,\alpha',
\gamma,\gamma')\\
&\leq C(s)\cdot \sum_{j\in \C{J} \setminus \{\gamma\}}\|
\chi_L^{+}{R}^{C_L}_FU_{\gamma'}^\frac12 \|^{3s}_{\HS}\cdot \|
U_j^\frac12\widehat{R}^{C_L} U_{\beta'}R^{C_L}_F\chi_y\|^{3s}_{\HS}
\end{align*}
so that, as the sum has only three terms,
\[
\expect( \|\widetilde{Q}_{\beta',\gamma'}\|^{3s}|\alpha,\alpha',
\gamma,\gamma')^\frac13 \leq C(s)\cdot \underbrace{\sum_{j\in \C{J}
\setminus \{\gamma\}}\| \chi_L^{+}{R}^{C_L}_FU_{\gamma'}^\frac12
\|^{s}_{\HS}\cdot \|
  U_j^\frac12\widehat{R}^{C_L}
U_{\beta'}R^{C_L}_F\chi_y \|^{s}_{\HS}}_{\Sigma_Q}.
\]
Similarly,
\[
\expect( \|  \widetilde{T}_{\beta,\gamma}\|^{3s} | \alpha,\alpha',
\gamma,\gamma')^\frac13 \leq C(s)\cdot \underbrace{\sum_{j\in \C{J}
\setminus \{\gamma'\}} \| \chi_xR^{B_L}_F U_\beta\widehat{R}^{B_L}
U_j^\frac12 \|^s_{\HS}\cdot \| U_ \gamma^\frac12 R^{B_L}_F\chi_L^{-}
\|^s_{\HS}}_{\Sigma_T} .
\]
Now $\Sigma_T$ and $\Sigma_Q$ are independent so that
\begin{equation} \label{eq:A4bound}
A^4_{\alpha,\alpha',\beta,\beta'\gamma,\gamma'}\leq C(s)
L^{2s(d-1)} \cdot \expecthat\expect[\Sigma_T]
\cdot\expecthat\expect[\Sigma_Q] \cdot e_{\alpha,\alpha'} .
\end{equation}
Since the $\xi_j$ and the $\widehat{\xi}_j$ have the same
distribution, we can omit the hats in $\widehat{R}^{C_L}$ and
$\widehat{R}^{B_L}$ and replace $\expecthat\expect$ by $\expect$
in the bounds for $\expecthat\expect[\Sigma_T]$ and
$\expecthat\expect[\Sigma_Q]$ to be derived below.

%------------------------%
\subsubsection*{Step 4}
We start with the $Q$-term. Combes-Thomas,
Proposition \ref{prop:CTHS} gives
\[
\|\chi_L^{+}R^{C_L}_F U_{\gamma'}^\frac12 \|^{s}_{\HS}\leq C
L^{s(d-1)} \ee^{-\mu_1 s ||\gamma'-x|-\frac{L}{2}|} .
\]
This will be used to deal with the term for $j=\gamma'$ which
appears in the sum over $\C{J}\setminus \{\gamma\}$; since $\| AB
\|_{\HS}\leq\| A\|\, \| B \|_{\HS}$ we get that:
\begin{align} \label{eq:gamma'bound}
&\| \chi_L^{+}{R}^{C_L}_FU_{\gamma'}^\frac12 \|^{s}_{\HS} \cdot
\|U_{\gamma'}^\frac12 R^{C_L}
U_{\beta'}R^{C_L}_F\chi_y \|^{s}_{\HS} \notag \\
&\qquad\quad
\leq
C L^{s(d-1)}\ee^{-\mu_1 s ||\gamma'-x|-\frac{L}{2}|} \|
  U_{\gamma'}^\frac12 {R}^{C_L}
U_{\beta'}^\frac12\|^{s}\cdot \| U_{\beta'}^\frac12
R^{C_L}_F\chi_y \|^{s}_{\HS} \notag \\
&\qquad\quad
\leq C
L^{s(d-1)}\sum_{\substack{x'\in Z(\gamma')\\
                                                y'\in Z(\beta')}}
\ee^{-\frac{c}{L}|x-x'|-\mu_1 s|y-y'|}\|
\chi_{x'}R^{C_L}\chi_{y'}\|^{s}.
\end{align}
For the terms $j=\alpha$ and $j=\alpha'$ in the sum we borrow from
$e_{\alpha,\alpha'}$ above and use that
\[
e_{\alpha,\alpha'}^\frac13\leq C\cdot\ee^{-c|x-x'|/L}
\]
if $j\in \{\alpha, \alpha'\}$ and $x'\in Z(j)$:
\begin{align} \label{eq:alal'bound}
&e_{\alpha,\alpha'}^\frac13\|
\chi_L^{+}{R}^{C_L}_FU_{\gamma'}^\frac12 \|^{s}
 \cdot \|
  U_j^\frac12 R^{C_L}
U_{\beta'}R^{C_L}_F\chi_y \|^{s}_{\HS} \notag \\
& \leq C L^{s(d-1)} \ee^{-\mu_1 s ||\gamma'-x|-\frac{L}{2}}
\sum_{\substack{x'\in Z(\gamma')\\
                                                y'\in Z(\beta')}}
\ee^{-\frac{c}{L}|x-x'|-\mu_1 s|y-y'|}\|
\chi_{x'}R^{C_L}\chi_{y'}\|^{s} .
\end{align}
Summing each of the three contributions from
\eqref{eq:gamma'bound} and \eqref{eq:alal'bound} to
$e_{\alpha,\alpha}^{\frac13} \Sigma_Q$ over $\beta', \gamma'$ (and
extending the $x'$-sum in \eqref{eq:alal'bound} to all of
$\D{Z}^d$) gives
\begin{equation}\label{Q}
\sum_{\beta',\gamma'\in\C{I}}e_{\alpha,\alpha'}^\frac13
\expecthat\expect[\Sigma_Q] \leq C
L^{2(d-1)}\sum_{x',y'\in\D{Z}^d} \ee^{-\frac{c}{L}|x-x'|-\mu_1 s
|y-y'|}\,\tau_{x',y'}.
\end{equation}

We now show that summation over $\alpha,\alpha',\beta,\beta'$
gives a small prefactor. By exponential decay:
\begin{equation}\label{S}
\sum_{\alpha,\alpha'}e_{\alpha,\alpha'}^\frac13\leq CL^{2d} .
\end{equation}

%------------------------%
\subsubsection*{Step 5}
We analyze
\[
\| \expect(\chi_xR^{B_L}_F U_\beta {R}^{B_L} U_j^\frac12 \|^s_{\HS})
\leq\| \chi_xR^{B_L}_F U_\beta^\frac12\|^s_{\HS}\cdot \expect(\|
U_\beta^\frac12 {R}^{B_L} U_j^\frac12\|^s) .
\]
If $|\beta-j|<\frac14 L$ then either $|x-\beta|\geq\frac18 L$ or
$\dist(j,\partial C_L)\geq  \frac18 L$. Since $j\in \C{J} \setminus
\{\gamma'\}$
\[
\text{either }\expect(\| \chi_xR^{B_L}_F U_\beta {R}^{B_L} U_j^\frac12 \|^s_{\HS}),
\quad e_{\alpha,\alpha'}^\frac13 \quad \text{or} \quad \|
U_\gamma^\frac12 R^{B_L}_F\chi_L^-\|^s_{\HS}
\]
is bounded by $L^{-\frac12 m_d}$; see Step 2 above. If, on the
other hand $|\beta-j|\ge\frac14 L$ we can use Lemma~\ref{startup}
above to estimate
\[
\expect(\| U_\beta^\frac12 {R}^{B_L} U_j^\frac12\|^s) \leq C\cdot
L^{-\frac12 m_d} .
\]
Summing up these terms we get that
\begin{equation}\label{T}
\sum_{\beta,\gamma\in\C{I}}e_{\alpha,\alpha'}^\frac13
\expecthat\expect[\Sigma_T] \leq C L^{3d-1-\frac12 m_d}
\end{equation}
since $\beta,\gamma$ run through at most $cL^d$ different points
of $\C{I}$ in $B_L$. Putting the estimates from
(\ref{Q}),(\ref{T}),(\ref{S}) together we arrive at:
\[
\sum_{\alpha,\alpha',\beta,\beta',\gamma,\gamma'}
A^4_{\alpha,\alpha',\beta,\beta',\gamma,\gamma'}\leq C\cdot
L^{9d-5-\frac12 m_d} \sum_{x',y'\in
\D{Z}^d}\ee^{-\frac{c}{L}|x-x'|-\mu_1 s |y-y'|}\,\tau_{x',y'}
\]
which is the desired bound. To deal with the other terms appearing
in $A_{\alpha,\alpha',\beta,\beta',\gamma,\gamma'}$ we just
combine the corresponding steps to control the $T$ and $Q$-sums
respectively.

This concludes the proof of Proposition~\ref{master}.
\end{proof}
%------------------------%

For energies sufficiently close to $E_0$ we will now complete the
proof of exponential decay of $\tau_{x,y}$, and thus of
Theorem~\ref{T1}, by applying a discrete Gronwall-type argument to
the recursion inequality established in Proposition~\ref{master}.

For $\mu>0$ consider the weighted $\ell^{\infty}$-space
\[
X=\ell^{\infty}(\D{Z}^{2d};\ee^{\mu\abs{x-y}/2}),
\]
i.e., for $\psi=(\psi_{x,y})$,
\[
\norm{\psi}_X=\sup_{x,y\in\D{Z}^d}\ee^{\mu\abs{x-y}/2}\abs{\psi_{x,y}}.
\]

%------------------------%
\begin{lem}    \label{lem.estimate.operator.A}
The operator $A$ defined by
\[
(A\psi)_{x,y}=\sum_{x',y'}
\ee^{-\mu(\abs{x-x'}+\abs{y-y'})}\psi_{x',y'}
\]
is bounded as an operator in $X$ as well as an operator in
$\ell^{\infty}(\D{Z}^{2d})$ with
\begin{equation}    \label{eq.estimate.operator.A.X.linfty}
\norm{A}_X\leq C(d)\mu^{-2d}\ \text{ and }\
\norm{A}_{\ell^{\infty}}\leq C(d)\mu^{-2d}.
\end{equation}
\end{lem}
%------------------------%

%------------------------%
\begin{proof}[Proof of Lemma \ref{lem.estimate.operator.A}]
The norm of $A$ in $X$ is the same as the norm of the operator
$\hat A$ in $\ell^{\infty}(\D{Z}^{2d})$ with kernel
\[
\hat A_{xyx'y'} = \ee^{\mu\abs{x-y}/2}
\ee^{-\mu(\abs{x-x'}+\abs{y-y'})} \ee^{-\mu\abs{x'-y'}/2}.
\]
Thus
\begin{align*}
\norm{A}_X=\norm{\hat A}_{\ell^{\infty}}
&=\sup_{x,y}\sum_{x',y'}\hat A_{xyx'y'}\\
&\leq
C\sup_{x,y} \iint\dd x'\dd y'\ee^{\mu\abs{x-y}/2}\ee^{-\mu(\abs{x-x'}+\abs{y-y'})}
\ee^{-\mu\abs{x'-y'}/2}\\
&= C\sup_{\Delta} \iint\dd s\,\dd p\,\ee^{\mu\abs{\Delta}/2}
\ee^{-\mu(\abs{s}+\abs{p-\Delta})-\mu\abs{p-s}/2},
\end{align*}
with the substitutions $s=x-x'$, $p=y'-x$, $\Delta=y-x$.

Bound the latter exponent through
\begin{align*}
\mu(\abs{s}+\abs{p-\Delta})
& +\frac{\mu}{2}\abs{p-s} \\
& =
\Bigl(\mu-\frac{\mu}{2}\Bigr)(\abs{s}+\abs{p-\Delta})
+\frac{\mu}{2}(\abs{\Delta-p}+\abs{p-s}+\abs{s})\\
&\geq
\frac{\mu}{2}(\abs{s}+\abs{p-\Delta})+\frac{\mu}{2}\abs{\Delta}.
\end{align*}
After cancellation the integral factorizes and gives
\eqref{eq.estimate.operator.A.X.linfty} for $\norm{A}_X$ after
scaling.
The bound for $\norm{A}_{\ell^{\infty}}$ is found more
directly.
\end{proof}
%--------------%

This may be applied to the situation of Proposition~\ref{master}
as it shows that for $L$ sufficiently large the operator $A$ with
kernel
\[
A_{xyx'y'}= L^{-2d-\kappa}\,\ee^{-c(|x-x'|+|y-y'|)/L}
\]
has norm less than one, both as an operator in
$X=\ell^{\infty}(\D{Z}^{2d};\ee^{c\abs{x-y}/2L})$ and an operator in
$\ell^{\infty}(\D{Z}^{2d})$. Fix this $L$ and choose $\delta =
L^{-m}$, $I=[E_0, E_0 + \delta]$ in Theorem~\ref{T1} and the
definition of $\tau_{x,y}$.

The recursion inequality \eqref{eq:recursion} now takes the form
\begin{equation}    \label{eq.simple.estimate.t.xy}
\tau_{x,y}\leq (A\tau)_{x,y}+b_{x,y}
\end{equation}
with $b_{x,y}:=C \ee^{-c\abs{x-y}/L}$. The conclusion of the proof
of Theorem~\ref{T1} is now the content of

%--------------%
\begin{lem} \label{lem.t.in.X}
\[
\tau=(\tau_{x,y})\in X.
\]
\end{lem}
%--------------%

%--------------%
\begin{proof}[Proof of Lemma~\ref{lem.t.in.X}]
With $\mu=\frac{c}{L}$ define the diagonal operator
\[
\C{D}=\diag(\ee^{\mu\abs{x-y}/2}),
\]
which is an isometry from $X$ to $\ell^{\infty}(\D{Z}^{2d})$. Let
$\hat \tau= \C{D}\tau$ and $\hat b=\C{D} b\in\ell^{\infty}$. Let
$\hat A=\C{D} A\C{D}^{-1}$. Then \eqref{eq.simple.estimate.t.xy}
implies that componentwise
\begin{equation}   \label{eq.simple.estimate.hat.t}
\hat\tau \leq\hat A\hat \tau+\hat b.
\end{equation}
Since $\tau=\C{D}^{-1}\hat\tau$ is bounded and $A$ a bounded
operator in $\ell^{\infty}(\D{Z}^{2d})$, we have that $\hat\tau
\in Y:=\ell^{\infty}(\D{Z}^{2d};\ee^{-\mu\abs{x-y}/2})$ and $\hat A$
is a bounded operator in $Y$ with non-negative kernel. Thus we
obtain from \eqref{eq.simple.estimate.hat.t} that
\[
\hat A^n\hat\tau \leq\hat A^{n+1}\hat\tau +\hat A^n\hat b
\]
holds with finite components. Summation yields
\[
\hat\tau \leq\hat A^{N+1}\hat\tau +\sum_{n=0}^N\hat A^n\hat b
\]
and thus
\[
\tau \leq A^{N+1}\tau +\sum_{n=0}^NA^nb
\]
for all $N$.

$A\colon\ell^{\infty}\to\ell^{\infty}$ is a contraction and $\tau
\in\ell^{\infty}$. Thus $A^{N+1}t\to 0$ in $\ell^{\infty}$ and
componentwise. Also, $A\colon X\to X$ is a contraction and $b\in
X$. Thus $\sum_{n=0}^NA^nb\to(I-A)^{-1}b\in X$ and componentwise
as $N\to\infty$. We conclude
\[
\tau \leq(I-A)^{-1}b\in X.
\]
Lemma~\ref{lem.t.in.X} is proved.
\end{proof}
%--------------%

%-------------------------------------------------------------%
\section{On the proof of Theorem~\ref{T2}}
\label{sec:proofT2}

That the localization properties stated in Theorem~\ref{T2} follow
from the fractional moment bound for the resolvent established in
Theorem~\ref{T1} was demonstrated in \cite{AENSS}. Here we want to
comment on two minor changes in the argument which are due to our
somewhat different set-up.

First we note that spectral and dynamical localization as
established in parts (a) and (b) of Theorem~\ref{T2} hold for
restrictions of $H$ to arbitrary open domains $G$, and, in
particular, that the exponential decay established in equation
\eqref{eq:dynloc} holds with respect to the standard distance
$|x-y|$ rather than the domain adapted distance $\dist_G(x,y)$ used
in \cite{AENSS}. Given that the corresponding bound \eqref{eq3} in
Theorem~\ref{T1} is true for arbitrary $G$ and in standard distance,
this follows with exactly the same proof as in Section~2 of
\cite{AENSS} (with one exception discussed below). That the authors
of \cite{AENSS} chose to work with the domain adapted distance was
in order to include more general regimes in which extended surface
states might exist. This is not the case in the regime considered
here.

Second, let us provide a few details on how to eliminate the use of
the covering condition \eqref{eq:cover} from the proof of
\eqref{eq:dynloc} provided in Section~2 of \cite{AENSS}. As done
there one first considers bounded open $\Lambda \subset \D{R}^d$ and
defines
\[
Y_{\Lambda}(I;x,y) := \sup_{f\in C_c(I),\,|f|\leq 1} \|\chi_x
f(H^{\Lambda}) \chi_y\|.
\]
If $E_n$ and $\psi_n$ are the eigenvalues and corresponding
eigenfunctions of $H^{\Lambda}$ and $f$ is as above, then
$f(H^{\Lambda}) = \sum_{n:\,E_n\in I} f(E_n) \scal{\psi_n,\,\cdot\,}\psi_n$ readily implies
\[
Y_{\Lambda}(I;x,y) \leq\sum_{n:\,E_n\in I} \|\chi_x \psi_n\|\cdot\|\chi_y
\psi_n\|.
\]
At this point we modify the argument of \cite{AENSS} and write
\begin{align*}
\chi_y \psi_n
&=
\chi_y (H_F^{\Lambda}-E_n)^{-1}
(H_F^{\Lambda}-E_n) \psi_n \\
&=
\chi_y (H_F^{\Lambda}-E_n)^{-1} W\psi_n \\
&=
\sum_{\alpha \in \C{I}} \xi_{\alpha} \chi_y
(H_F^{\Lambda}-E_n)^{-1} U_{\alpha} \psi_n .
\end{align*}
As all $E_n\in I$ have a uniform distance from $\inf
\sigma(H_F^{\Lambda})$ we get from Combes-Thomas
Proposition~\ref{prop:CT} that
\begin{align*}
\|\chi_y \psi_n\|
&\leq
C \sum_{\alpha} \|\chi_y
(H_F^{\Lambda}-E_n)^{-1} U_{\alpha}^{1/2}\|\cdot\|U_{\alpha}^{1/2}
\psi_n\| \\
&\leq
C \sum_{\alpha} \ee^{-\mu_0 |y-\alpha|} \|U_{\alpha}^{1/2}
\psi_n\| .
\end{align*}
Inserting above yields
\[
Y_{\Lambda}(I;x,y) \leq C\sum_{\alpha} \ee^{-\mu_0 |y-\alpha|}
Q_1(I;x,\alpha)
\]
with $Q_1(I;x,\alpha) = \sum_{n:\,E_n \in I} \|\chi_x \psi_n\|\cdot
\|U_{\alpha}^{1/2} \psi_n\|$ defined as in \cite{AENSS}, where the
bound $\expect(Q_1(I;x,\alpha)) \leq C\ee^{-\mu_1|x-\alpha|}$ is
established without any further references to the covering
condition. Thus we conclude
\begin{equation} \label{eq:finitedynloc}
\expect\bigl(Y_{\Lambda}(I;x,y)\bigr) \leq C \ee^{-\mu_2|x-y|} .
\end{equation}
The rest of the proof of Theorem~\ref{T2}, in particular the
extension of \eqref{eq:finitedynloc} to infinite volume and a
supremum over arbitrary Borel functions, follows the argument in
\cite{AENSS} without change.

%-------------------------------------------------------------%
\section{Localization for continuum random surface models}
\label{sec:surface}

Random surface models have attracted quite some interest with most
of the work dealing with the discrete case \cite{BS:98, G:95, JL,
JM:98, JM:99a, Jaksic/Molchanov, JMP-98} and some with the
continuum case \cite{HK, KW, BdMS, BSS}, as we do here. Our aim in
this section is to show that under suitable conditions such
surface models obey condition (A4) above. To achieve it, we
combine recent results from \cite{KW} with a technique from
\cite{StoMPAG}.

As usual, the background is assumed to be partially periodic:
\begin{enumerate}[{(B}1)]
\item
Fix $1\leq d_1 \leq d $ and write
$\D{R}^d=\D{R}^{d_1}\times \D{R}^{d_2}$, $x=(x_1,x_2)$; assume that
$V_0\in L^2_{\text{loc,unif}}(\D{R}^d)$ is real-valued and periodic with
respect to the first variable, i.e.,
\[
V_0(x_1+m,x_2)=V_0(x_1,x_2)\text{  for  }m\in \D{Z}^{d_1}.
\]
Denote $H_0:=-\Delta +V_0$.
\end{enumerate}
In order to state our second requirement, let us recall some facts
from Bloch theory. For more details, see \cite{KW}. For $V_0, H_0$ as in
(B1) we get a direct integral decomposition
\[
H_0= (2\pi)^{-d_1} \int_{\D{T}^{d_1}}^\oplus h_\theta\,\dd\theta,
\]
where $\D{T}^{d_1} = \D{R}^{d_1}/(2\pi \D{Z})^{d_1}$ is the
$d_1$-dimensional torus and
\[
h_\theta=-\Delta +V_0 \text{  in  }L^2(S_1)
\]
with $\theta$-periodic boundary conditions on the unit strip $S_1
= \Lambda_1(0) \times \D{R}^{d_2}$. We now fix the assumption
\begin{enumerate}[{(B}1)]
\addtocounter{enumi}{1}
\item
\[
\inf\sigma(h_0) < \inf\sigma_{\text{ess}}(h_0)  .
\]
\end{enumerate}
It is well known that under (B2) we have that
\[
E_0:=\inf\sigma(H_0)=\inf\sigma(h_0)
\]
and there is a positive
eigensolution $\psi_0$ of the distributional equation
\[
H_0\psi_0=E_0\psi_0,
\]
see \cite{KW, KW1} and the references therein.
Finally, our random perturbation is assumed to satisfy
\begin{enumerate}[{(B}1)]
\addtocounter{enumi}{2}
\item
The set $\C{I}\subset\D{R}^d$, where the random impurities
   are located, is uniformly discrete, i.e., $\inf\{\abs{\alpha-\beta}:
   \alpha\not= \beta\in \C{I}\}=:r_\C{I}>0$.
Moreover $\C{I}$ is dense near the surface $\D{R}^{d_1}\times \{ 0\}$
in the sense that there exist $R_\perp, c_\perp>0$ such that for
$L$ large enough and $x_1\in\D{R}^{d_1}$:
\[
\#\bigl[\C{I}\cap\bigl(\Lambda_L(x_1)\times \Lambda_{R_\perp\!}(0)\bigr)\bigr]\geq
c_\perp L^{d_1} .
\]
\end{enumerate}
We will see that (B1)-(B3) ensure (A4) from Section 1. Of course, there
might be other ways to verify (A4) for surface-like potentials so that
Theorems 1 and 2 could, in principle, be used for other examples.

%------------------------%
\begin{thm}\label{T3}
Assume \emph{(B1)-(B3)} and \emph{(A3)}.  Then there exist $\delta>0$, $0<s<1$,
   $\mu>0$ and $C<\infty$ such that for $I:=\croch{E_0,E_0+\delta}$, all
   open sets $G \subset \D{R}^d$ and $x,y\in \D{R}^d$
\begin{equation}
\sup_{E\in I,\,\varepsilon >0}\expect (\norm{\chi_{x}
(H^G-E-\ii\varepsilon)^{-1}\chi_{y}}^{s}) \leq C\,\ee^{-\mu
\abs{x-y}}.
\end{equation}
In particular, the following consequences hold:
\begin{enumerate}[\rm(a)]
\item
The spectrum of $H^G$ in $I$ is almost surely pure point with
   exponentially decaying eigenfunctions.
\item
There are $\mu >0$ and $C<\infty$ such that for all
   $x,y\in\D{Z}^d$,
\begin{equation}
\expect\bigl(\sup_{t\in\D{R}}\norm{\chi_{x} \ee^{-\ii
tH^G}P_I(H^G)\chi_y} \bigr) \leq C\ee^{-\mu\abs{x-y}}.
\end{equation}
\end{enumerate}
\end{thm}
%------------------------%

The rest of this section is devoted to deducing (A4) under the
assumptions of the Theorem. Note that this will be accomplished
once we have shown the following, where
\[
S_L=S_L(x_1):= \Lambda_L(x_1)\times\D{R}^{d_2}
\]
denotes the strip of side length $L$ centered at $x_1\in
\D{R}^{d_1}$ perpendicular to the ``surface'' $\D{R}^{d_1}\times
\{ 0\}$.

%------------------------%
\begin{prop}\label{p3}
For all $\gamma, \xi >0$ there exists $L(\gamma, \xi)$ such that
for all odd integers $L\geq L(\gamma, \xi)$ and $x_1 \in
\D{Z}^{d_1}$:
\begin{equation} \label{eq:ILSbound}
\prob\bigl\lbrace\sigma(H^{S_L(x_1)})\cap [E_0,E_0+L^{-\gamma}]\not= \emptyset\bigr\rbrace
\leq L^{-\xi} .
\end{equation}
\end{prop}
%------------------------%

In fact, (A4)(ii) then follows, since $H^{\Lambda_L(x)}\ge
H^{S_L(x_1)}$ and therefore $E_0 \leq\inf  \sigma(H^{S_L(x_1)})\le
\inf \sigma(H^{\Lambda_L(x)})$.

We will actually prove the analogue of Proposition \ref{p3} with
Dirichlet boundary conditions replaced by suitable Robin boundary
conditions that are defined using the periodic ground state
$\psi_0$ introduced above. Assume, for later convenience, that
\[
\int_{S_1} |\psi_0(x)|^2\dd x=1.
\]
We consider on $S_L$, $L\in
2\D{N}-1$, \emph{Mezincescu boundary conditions}, given as
follows. Let
\[
\chi(x):= -\frac{1}{\psi_0(x)}\nabla_n\psi_0(x),
\]
where $\nabla_n$ denotes the outer normal derivatives. The
Mezincescu
boundary condition can be thought of as the following requirement
for functions $\phi$ in the domain of $H^{S_L}_\chi$:
\[
\nabla_n\phi(x)=-\chi(x)\phi(x)\text{  for  }x\in
\partial S_L .
\]
For
the formal definition of $H^{S_L}_\chi$ via quadratic forms and more
background, see Mezincescu's original paper \cite{M} as well
as \cite{KW1, KW}. In particular, we immediately get
the following important relations in the sense of the corresponding
quadratic forms:
\begin{equation}\label{eq31}
H^{S_L}_\chi \leq H^{S_L}
\end{equation}
as well as
\begin{equation}\label{eq32}
H^{S_L}_\chi \geq\bigoplus_{k=1}^n H^{S_{l_k}(y_k)}_\chi ,
\end{equation}
whenever the strip $S_L$ is divided into disjoint strips
$S_{l_k}(y_k)$ whose closures exhaust the closure of $S_L$.

%------------------------%
\begin{proof}[Proof of Proposition \ref{p3}]
Due to the form inequality (\ref{eq31}) above it remains to prove
the estimate for $H^{S_L}_\chi$.

Denoting the bottom eigenvalue of an operator $H$ by $E_1(H)$
(\textit{caution}\/: here our notation differs from the one in \cite{KW1,
KW}, where the second eigenvalue is denoted by $E_1(H)$) we see
that
\[
\sigma(H^{S_L}_\chi)\cap [E_0,E_0+L^{-\gamma}]\not= \emptyset
\Longleftrightarrow
E_1(H^{S_L}_\chi)\leq E_0+ L^{-\gamma} .
\]

%------------------------%
\subsubsection*{Step 1}
There exist $b,K, \beta>0$ such that
\begin{equation}
\label{eq:33}
  \prob\bigl\lbrace E_1(H^{S_L}_\chi)\leq E_0+ bL^{-2}\bigr\rbrace\leq K\cdot
  \exp(-K\cdot L^{d_1}) .
\end{equation}
We use here the method from \cite{StoMPAG}. Denote $H(t,\omega):=
(H_0+tV_\omega)^{S_L}_\chi$, and its first eigenvalue by
$E_1(t,\omega)$. Since $E_1(t,\omega)$ increases in $t$ the event
in (\ref{eq:33}) implies that $E_1(t,\omega)$ be small for all
$t\leq 1$ which in turn implies that $E_1'(0,\omega)$ must be
small.

We infer from \cite{KW}, Theorem~3.25 that the gap between the
first two eigenvalues satisfies
\[
E_2(0,\omega) -E_1(0,\omega)
\geq\const L^{-2}.
\]
As in \cite{StoMPAG}, Lemma 2.3 this gives that
\begin{equation}                 \label{eq:*}
| E_1(t,\omega) - (E_0+t\cdot E_1'(0,\omega))|\le
KL^2\cdot t^2 \text{  for  }0\leq t\ \le\tau\cdot L^{-2} .
\end{equation}
Now assume that
\[
E_1(H^{S_L}_\chi)\leq E_0+ bL^{-2}
\]
for $b>0$. From (\ref{eq:*}) we get that
\[
E_1'(0,\omega)\leq c(b)
\]
with $c(b)\to 0$ for $b\to 0$.

On the other hand
\[
E_1'(0,\omega)= (V_\omega\psi_{0,L}|\psi_{0,L})
\]
where $\psi_{0,L}$ is the normalized ground state of
$H^{S_L}_{0,\chi}$. Now, the boundary condition of
$H^{S_L}_{0,\chi}$ is defined so as to make sure that $\psi_{0}$
is an eigenfunction; see the discussion in \cite{KW}. Therefore
$\psi_{0,L}=L^{-d_1/2}\psi_{0}$ and we get
\begin{align*}
E_1'(0,\omega)
&=
(V_\omega\psi_{0,L}|\psi_{0,L}) \\
&=
L^{-d_1}\sum_{\alpha\in \C{I}} \eta_\alpha(\omega)\cdot
\int_{S_L}U_\alpha(x)|\psi_0(x)|^2\dd x\\
&\geq
L^{-d_1}\sum_{\alpha\in \C{I}\cap S_{L-r_U}} \eta_\alpha(\omega)\cdot
c_U\cdot\int_{\Lambda_{r_U}(\alpha)}|\psi_0(x)|^2\dd x .
\end{align*}
Since, by (B3), there are at least $c_\perp (L-r_U)^{d_1}$
elements of $ \C{I}\cap S_{L-r_U}$ in $\D{R}^{d_1}\times
\Lambda_{R_\perp\!}(0)$ and
\[
\inf_{(x_1,x_2)\in\D{R}^{d_1}\times \Lambda_{R_\perp\!}(0)}
\int_{\Lambda_{r_U}(x_1,x_2)}|\psi_0(x)|^2\dd x >0
\]
we arrive at
\begin{equation}
E_1'(0,\omega) \geq c_1\cdot \frac{1}{|\C{I}_\perp|} \sum_
{\alpha\in \C{I}_\perp}\eta_\alpha(\omega)
\end{equation}
with $c_1>0$ and independent variables $\eta_\alpha$ running
through an index set $\C{I}_\perp$ of cardinality at least $c_2
L^{d_1}$. If we now choose $b>0$ so small that $\frac{c(b)}{c_1}<
M$ where $M$ is smaller than the mean of all the $\eta_\alpha$'s
we get that:
\begin{align*}
\prob\bigl\lbrace  E_1(H^{S_L}_\chi)\leq E_0+ bL^{-2}\bigr\rbrace
& \leq\prob\bigl\lbrace
c_1\cdot \frac{1}{|\C{I}_\perp|}
\sum_ {\alpha\in \C{I}_\perp}\eta_\alpha(\omega)\leq c(b) \bigr\rbrace\\
&\leq K\cdot \exp (-\beta_0 |\C{I}_\perp|)\\
&\leq K\cdot \exp (-\beta L^{d_1}),
\end{align*}
by a standard large deviation estimate; see \cite{Ka} or
\cite{Tal}, Theorem~1.4. This finishes the proof of Step 1.

%------------------------%
\subsubsection*{Step 2}
To deduce the desired bound from Step 1 we divide the strip $S_L$
into disjoint strips $S_{l_k}(y_k)$ whose closures exhaust the
closure of $S_L$ and such that
\[
L^{-\gamma}\leq b\cdot l_k^{-2}\leq 42 \cdot L^{-\gamma},\quad
 l_k\in 2\D{N}+1
\]
which is possible for $L$ large enough.

Their number $n$ is at
most $\const L^{(1-\frac{\gamma}{2})d_1}$. By (\ref{eq32}) we
know that
\[
  E_1\bigl(H^{S_L}_\chi\bigr)\geq\min_{1\leq k\leq n} E_1\bigl(H^{S_{l_k}(y_k)}_\chi\bigr)
\]
so that
\begin{align*}
\prob\bigl\lbrace  E_1(H^{S_L}_\chi)\leq E_0+ L^{-\gamma}\bigr\rbrace
&\leq
\prob\bigl\lbrace\min_{1\leq k\leq n} E_1(H^{S_{l_k}(y_k)}_\chi) \leq E_0+ L^{-\gamma}\bigr\rbrace\\
&\leq
\sum_{k=1}^n \prob\bigl\lbrace E_1(H^{S_{l_k}(y_k)}_\chi) \leq E_0+
L^{-\gamma}\bigr\rbrace\\
&\leq
\sum_{k=1}^n \prob\bigl\lbrace E_1(H^{S_{l_k}(y_k)}_\chi) \leq E_0+
b\cdot l_k^{-2}\bigr\rbrace\\
&\leq
n\cdot K\cdot  \exp (-\beta {l_k}^{d_1})\\
&\leq
L^{-\xi}
\end{align*}
provided $L$ is large enough.
\end{proof}
%------------------------%

%------------------------%
\begin{rems*}
(1)
In cases where the operator $H$ is ergodic, a stronger bound
than \eqref{eq:ILSbound} is provided in
\cite[Proposition~5.2]{KW}. Their bound is in terms of the
integrated density of states for which \cite{KW} establishes
Lifshits asymptotics. As we are only interested in localization
properties here, the bound \eqref{eq:ILSbound} suffices and allows
to handle the non-ergodic random potentials defined in (B3) and
(A3).

(2)
We have established localization near the bottom of the
spectrum for the random surface models considered in this section.
If $d_1=1$ one expects for physical reasons that the entire
spectrum of $H$ below $\inf\sigma(H_F)$ (see (A4)) is localized.
A corresponding result for lattice operators has been proven in
\cite{Jaksic/Molchanov} (in situations where $H_F$ is the discrete
Laplacian and $d_2=1$). To show this for continuum models remains
an open problem.
\end{rems*}
%------------------------%

%-------------------------------------------------------------%
\section{Anderson models with displacement}
\label{sec:displacement}

By considering the special case $d_1=d$, the results of the
previous Section also cover ``usual'' Anderson models, sometimes
also called alloy models. Note that in this case (B2) becomes
trivial. Let us nevertheless state the assumptions and result
again for this case, mainly because we want to point out below
that the obtained bounds hold uniformly in the geometric
parameters describing the random potential. This will then be
applied to models with random displacements. Here are the
assumptions we rely upon:
\begin{enumerate}[{(D}1)]
\item
$V_0\in L^2_{\text{loc,unif}}(\D{R}^d)$ is real-valued and periodic.
\item
The set $\C{I}\subset\D{R}^d$, where the random impurities
   are located, is uniformly discrete, i.e., $\inf\{\abs{\alpha-\beta}:
   \alpha\not= \beta\in \C{I}\}=:r_\C{I}>0$ and uniformly dense, i.e.,
there exists $R_\C{I}>0$ such that $\Lambda_{R_\C{I}}(x)\cap\C{I}
\not =\emptyset$ for every $x\in\D{R}^d$.
\end{enumerate}

%------------------------%
\begin{thm}\label{T10}
Assume \emph{(D1), (D2)} and \emph{(A3)}.  Then there exist $\delta>0$, $0<s<1$,
   $\mu>0$ and $C<\infty$ such that for $I:=\croch{E_0,E_0+\delta}$, all
   open sets $G \subset \D{R}^d$ and $x,y\in \D{R}^d$
\begin{equation}
\sup_{E\in I,\varepsilon >0}\expect (\norm{\chi_{x}
(H^G-E-\ii\varepsilon)^{-1}\chi_{y}}^{s}) \leq C\,\ee^{-\mu
\abs{x-y}}.
\end{equation}
In particular, the following consequences hold:
\begin{enumerate}[\rm(a)]
\item
The spectrum of $H^G$ in $I$ is almost surely pure point with
   exponentially decaying eigenfunctions.
\item
There are $\mu_1 >0$ and $C_1<\infty$ such that for all
   $x,y\in\D{Z}^d$,
\begin{equation}
\expect\bigl(\sup_{t\in\D{R}}\norm{\chi_{x} \ee^{-\ii
tH^G}P_I(H^G)\chi_y} \bigr) \leq C_1\ee^{-\mu_1\abs{x-y}}.
\end{equation}
\end{enumerate}
Here all the constants $\delta, s, C, \mu, C_1, \mu_1$ can be
chosen to only depend on the potential through the parameters
$V_0,\eta_{\max},M_\rho,c_U,C_U,r_U,R_U,r_\C{I},R_\C{I}$.
\end{thm}
%------------------------%

To this end we first observe that (D1), (D2) and (A3) imply (A4)
with constants $E_F$, $m$ and $L^*$ only depending on the listed
parameters:

%------------------------%
\begin{prop}\label{p11}
Assume \emph{(D1), (D2)} and \emph{(A3)}. Then there exist
\begin{align*}
&E_1
= E_1(V_0,\eta_{\max},M_\rho,c_U,C_U,r_U,R_U,r_\C{I},R_\C{I})>E_0 ,\\
&m
=m(V_0,\eta_{\max},M_\rho,c_U,C_U,r_U,R_U,r_\C{I},R_\C{I})\in
(0,2)
\end{align*}
and $L^*=L^*(\ldots)$ such that
\begin{enumerate}[\rm(1)]
\item
$E_F\geq E_1$.
\item
For $m_d:=42\cdot d$, all $L\geq L^*$ and $ x\in \D{Z}^d$:
\[
\D{P}\bigl(
     \sigma(H^{\Lambda_L(x)}(\omega))\cap[E_0,E_0+L^{-m}]
     \neq\emptyset\bigr)\leq L^{-m_d} .
\]
\end{enumerate}
\end{prop}
%------------------------%

%------------------------%
\begin{proof}
First we show that (D2) implies that there exist
$c_\C{I},C_\C{I}$ and $L_\C{I}$ depending only on
$r_\C{I},R_\C{I}$ such that for all $L\geq L_\C{I}$:
\begin{equation} \label{vol}
c_\C{I}\cdot  L^d\leq\#\left(\C{I}\cap \Lambda_L(x)\right)\leq C_\C{I}
\cdot   L^d .
\end{equation}
The upper bound follows from uniform discreteness:
\[
\#\left(\C{I}\cap \Lambda_L(x)\right)\cdot |B_{r_\C{I}/2}|
\leq |  \Lambda_{L+r_{\C{I}/2}}|\leq (2L)^d ,
\]
provided $L\geq r_\C{I}/2$. For the lower bound use uniform
denseness: Divide $\Lambda_L(x)$ into disjoint boxes of side length
$R_\C{I}$. If $L\geq 2R_\C{I}$ there are at least $(L/ 2{R_\C{I}})^d$
of them each of which contains at least one point from $\C{I}$.

Now we can use the analysis of the preceding
Section. Since the relevant quantities depend only on the
indicated parameters, the assertions follow.
\end{proof}
%------------------------%

With this uniform version of (A4) and the proofs provided in
Sections~\ref{sec:mainproofs} and \ref{sec:proofT2} we also get
corresponding uniform versions of Theorems~\ref{T1} and \ref{T2},
i.e.\ Theorem~\ref{T10}.

As a specific application of the previous observation, we can
start from an Anderson model as above and additionally vary the
set $\C{I}$ in a random way, as long as $r_\C{I}$ and $R_\C{I}$
obey uniform upper and lower bounds. Instead of formulating the
most general result in this direction we look at models that were
introduced in \cite{CH} and further studied in \cite{Ze}.
\begin{enumerate}[{(D}1)]
\addtocounter{enumi}{2}
\item
Let $\eta_j$, $j\in\D{Z}^d$ be independent random couplings, defined
on a probability space $\Omega$ with distribution $\rho_j$
and $U_j$ as in (A3).
\item
Let $x_j$, $j\in \D{Z}^d$ be independent random vectors
of length at most $\frac13$ in $\D{R}^d$; denote the corresponding
probability space by $\widetilde{\Omega}$.
\end{enumerate}
Define
\[
H(\omega,\widetilde{\omega}):=-\Delta+V_0+\sum_{j\in\D{Z}^d}
\eta_j(\omega)U_j(\,\cdot-j-x_j(\widetilde{\omega})) .
\]

%------------------------%
\begin{cor}
Assume \emph{(D1), (D3), (D4)}. Then, for $
H(\omega,\widetilde{\omega})$ as above there
  exist $\delta>0$, $0<s<1$,
   $\mu>0$ and $C<\infty$ such that for $I:=\croch{E_0,E_0+\delta}$, all
   open sets $G \subset \D{R}^d$ and $x,y\in \D{R}^d$
\begin{equation}
\sup_{E\in I,\varepsilon >0}\expecttilde
\expect (\norm{\chi_{x}
(H^G-E-\ii\varepsilon)^{-1}\chi_{y}}^{s}) \leq C\,\ee^{-\mu
\abs{x-y}}.
\end{equation}
In particular, the following consequences hold:
\begin{enumerate}[\rm(a)]
\item
The spectrum of $H^G$ in $I$ is almost surely pure point with
   exponentially decaying eigenfunctions.
\item
There are $\mu >0$ and $C<\infty$ such that for all
   $x,y\in\D{Z}^d$,
\begin{equation}
\expecttilde\expect\bigl(\sup_{t\in\D{R}}\norm{\chi_{x}
\ee^{-\ii tH^G}P_I(H^G)\chi_y} \bigr) \leq C\ee^{-\mu\abs{x-y}}.
\end{equation}
\end{enumerate}
\end{cor}
%------------------------%

%------------------------%
\begin{proof}
The corresponding inequality holds uniformly in $\widetilde{\omega}$
by what we proved above.
\end{proof}
%------------------------%

Note that in this last Corollary we have not assumed that the random
perturbations cover the whole space. In that respect our result
provides substantial progress as compared to \cite{CH,Ze}.

%-------------------------------------------------------------%
\begin{appendix}
%-------------------------------------------------------------%
\section{Some technical tools}

Here we collect some technical background which was used in
Section~2 above. All of this is known. We either provide
references or, for convenience, in some cases sketch the proof.

%-------------------------------------------------------------%
\subsection{Combes-Thomas bounds}
\label{CT}
Proofs of the following improved Combes-Thomas bound can be found
in \cite{BCH} (where it was first observed) and \cite{Stollmann}.
We state it here under assumptions which are sufficient for our
applications. In particular, we assume $d\leq 3$, while the result
holds in arbitrary dimension for a suitably modified class of
potentials. As above, for an open $G\subset \D{R}^d$ we denote by
$H^{G}$ the restriction of $-\Delta+V$ to $L^2(G)$ with Dirichlet
boundary conditions.

%------------------------%
\begin{prop} \label{prop:CT}
Let $d\leq 3$, $V\in L^2_{\textup{loc,unif}}(\D{R}^d)$ with $\sup_x \|
V\chi_{\Lambda_1(x)}\|_2 \leq M$. Let $M\geq 1$ and $R>0$. Then
there exist $c_1= c_1(M,R)$ and $c_2= c_2(M,R)$ such that the
following conditions
\begin{enumerate}[\rm(i)]
\item
$G\subset \D{R}^d$ open, $A,B \subset G$, $\dist(A,B) =:
\delta
>0$,
\item
$(r,s) \subset \rho(H^{G}) \cap (-R,R)$, $E\in (r,s)$ and
$\eta := \dist(E,(r,s)^c)>0$,
\end{enumerate}
imply the estimate
\begin{equation} \label{eq:CTbound}
\sup_{\varepsilon \in \D{R}} \|\chi_A (H^{G}-E-\ii\varepsilon)^{-1}
\chi_B \| \leq\frac{c_1}{\eta} \ee^{-c_2 \sqrt{s-r} \eta^{1/2}
\delta}.
\end{equation}
\end{prop}
%------------------------%

Note that the results in \cite{BCH} and \cite{Stollmann} are
stated for $\varepsilon =0$, but the proofs are easily adjusted to
show that the bounds are uniform in the additional imaginary part.

%-------------------------------------------------------------%
\subsection{Combes-Thomas bounds
                         in Hilbert-Schmidt norm}

A consequence of \eqref{eq:CTbound} is that $\|\chi_x
(H^{(G)}-E-\ii\varepsilon)^{-1} \chi_y\|$ decays exponentially in
$|x-y|$. Due to the restriction to $d\leq 3$ this is also true in
Hilbert-Schmidt norm:

%------------------------%
\begin{prop} \label{prop:CTHS}
Let $d\leq 3$, $V\in L^2_{\textup{loc,unif}}(\D{R}^d)$, $H=-\Delta + V$ in
$L^2(\D{R}^d)$ and $I\subset (-\infty, \inf \sigma(H))$ a compact
interval. Then there exist $C<\infty$ and $\mu>0$ such that
\begin{equation} \label{eq:HSCTbound}
\sup_{\substack{E\in I, \, \varepsilon>0\\
                              G\subset \D{R}^d\text{open}}}
                              \|\chi_x (H^{G}-E-\ii\varepsilon)^{-1}
\chi_y\|_{\HS} \leq C \ee^{-\mu |x-y|}
\end{equation}
for all $x,y \in \D{R}^d$.
\end{prop}
%------------------------%

%------------------------%
\begin{proof}
Let us sketch the proof by
combining several well known facts. To this end, let $S_p$ denote
the $p$-th Schatten class, i.e.\ the set of all bounded operators
$A$ such that $\|A\|_p := (\tr |A|^p)^{1/p} < \infty$. As $d\le
3$, by Theorem~B.9.3 of \cite{Simon:semi} we have
\begin{equation} \label{eq:simonbound}
\|\chi_x (H-E)^{-1/2}\|_p \leq C_1 < \infty
\end{equation}
for each $p>3$ and $E< \inf \sigma(H)$. The proof provided in
\cite{Simon:semi} shows that $C_1$ can be chosen uniform in $x\in
\D{R}^d$ and $E\in I$. In the sense of quadratic forms it holds
that $H^G \geq H$ for each open $G\subset \D{R}^d$, i.e.\
$\|(H-E)^{1/2}(H^G-E)^{-1/2}\| \leq 1$ for all $E< \inf \sigma(H)$,
see e.g.\ Section~VI.2 of \cite{Kato}. Thus
\begin{align} \label{eq:subsetbound}
\|\chi_x (H^G-E)^{-1/2} \|_p
&\leq\|\chi_x (H-E)^{-1/2}\|_p\,
\|(H-E)^{1/2} (H^G-E)^{-1/2}\| \notag\\
&\leq C_1<\infty.
\end{align}
The H\"older property of Schatten classes implies
\begin{equation} \label{eq:holderschatten}
\|\chi_x (H^G-E)^{-1} \chi_y \|_{p/2} \leq C_1^2
\end{equation}
uniformly in $x,y \in \D{R}^d$, $E\in I$ and $G\subset \D{R}^d$
open. From the resolvent identity
\[ \chi_x(H^G-E-\ii\varepsilon)^{-1}\chi_y = \chi_x(H^G-E)^{-1}\chi_y
+\ii\varepsilon \chi_x (H^G-E-\ii\varepsilon)^{-1} (H^G-E)^{-1} \chi_y
\]
we easily see that
\begin{equation} \label{eq:imagpart}
\|\chi_x (H^G-E-\ii\varepsilon)^{-1} \chi_y\|_{p/2} \leq C_2 < \infty
\end{equation}
holds uniformly also in the additional parameter $\varepsilon \in
\D{R}$. By Proposition~\ref{prop:CT} we also have $C_3<\infty$ and
$\mu_1>0$ such that
\begin{equation} \label{eq:normdecay}
\|\chi_x (H^G-E-\ii\varepsilon)^{-1} \chi_y\| \leq C_3 \ee^{-\mu_1
|x-y|}
\end{equation}
uniform in $G$, $E\in I$ and $\varepsilon \in \D{R}$. As we may
choose $p/2 \in (3/2,2)$, \eqref{eq:HSCTbound} follows from
\eqref{eq:imagpart} and \eqref{eq:normdecay} by interpolation,
more precisely from the fact that $\norm{\,\cdot\,}_{\HS} = \norm{\,\cdot\,}_2$
and $\|A\|_2^2 = \tr |A|^2 = \tr (|A|^{p/2} |A|^{2-p/2}) \le
\|A\|^{2-p/2} \|A\|_{p/2}^{p/2}$.
\end{proof}
%------------------------%

%-------------------------------------------------------------%
\subsection{A fractional-moment bound}

The next result and its proof are found in \cite{AENSS}, where it
played a central role in the extension of the fractional-moment
method to Anderson-type random Schr\"odinger operators in the
continuum.

Recall that an operator $A$ is called dissipative if
$\Im \scal{A\varphi, \varphi}\geq 0$ for all $\varphi \in D(A)$. It is
called maximally dissipative if it has no proper dissipative
extension. Below we also use the notation $|\cdot|$ for Lebesgue
measure in $\D{R}^2$.

%------------------------%
\begin{prop} \label{prop:weakbound}
There exists a universal constant $C<\infty$ such that for every
separable Hilbert space $\C{H}$, every maximally dissipative
operator $A$ in $\C{H}$ with strictly positive imaginary part
(i.e.\ $\Im \scal{A\varphi, \varphi}\geq\delta
\|\varphi\|^2$ for some $\delta>0$ and all $\varphi \in D(A)$),
for arbitrary Hilbert-Schmidt operators $M_1$, $M_2$ in $\C{H}$,
for arbitrary bounded non-negative operators $U_1$, $U_2$ in
$\C{H}$, and for all $t>0$ the following holds:
\begin{align} \label{eq:weakbound}
\bigl\lvert\bigl\{ (v_1,v_2) \in [0,1]^2:
& \:\|M_1 U_1^{1/2}
(A-v_1U_1- v_2U_2)^{-1} U_2^{1/2} M_2\|_{\HS} >t \bigr\}\bigr\rvert\notag\\
&\leq
C \|M_1\|_{\HS} \|M_2\|_{\HS} \cdot \frac{1}{t}.
\end{align}
\end{prop}
%------------------------%

The weak-$L_1$-type bound \eqref{eq:weakbound} yields a fractional
moment bound:

%------------------------%
\begin{cor} \label{cor:fracmoments}
Let $s\in (0,1)$. Then for the constant $C$ and operators $A$, $M_1$,
$M_2$, $U_1$, $U_2$ as in Proposition~\ref{prop:weakbound},
\begin{align} \label{eq:fracbound}
\int_0^1\dd v_1 \int_0^1\dd v_2
 \|M_1 U_1^{1/2}
&(A-v_1U_1-v_2U_2)^{-1} U_2^{1/2} M_2\|_{\HS}^s \notag \\
& \leq\frac{C^s}{1-s} \|M_1\|_{\HS}^s \|M_2\|_{\HS}^s.
\end{align}
\end{cor}
%------------------------%

This follows with layer-cake integration, which gives for the
l.h.s.\ of \eqref{eq:fracbound}
\[
\int_0^1\dd v_1 \int_0^1\dd v_2\norm{\ldots}^s
\leq\int_0^{\infty} \bigl\lvert\accol{(v_1,v_2)\in
[0,1]^2:\: \norm{\ldots}> t^{1/s}}\bigr\rvert\,\dd t.
\]
The integrand is bounded by the minimum of $1$ and a bound
following from \eqref{eq:weakbound}. Splitting the integral
accordingly leads to \eqref{eq:fracbound}.

%------------------------%
\begin{rems*}
(1)
The use of the interval $[0,1]$ as support of
$v_1, v_2$ in Proposition~\ref{prop:weakbound} and
Corollary~\ref{cor:fracmoments} is not essential. Using shifting
and scaling  it can be replaced by an arbitrary compact interval
$K$, with constants becoming $K$-dependent.

(2) In our applications maximally dissipative operators arise in
the form $A= -(S-E-\ii\varepsilon)$ for self-adjoint operators $S$,
with $\varepsilon>0$ providing a strictly positive imaginary part.

(3) Note that, as seen from the argument in \cite{AENSS}, a bound
like \eqref{eq:fracbound} also holds in the ``diagonal'' case,
i.e.\ for $\int_0^1\dd v\,\|M U^{1/2} (A-vU)^{-1} U^{1/2}
M\|_{\HS}^s$.
\end{rems*}
%------------------------%
\end{appendix}
%-------------------------------------------------------------%
\subsection*{Acknowledgement}
S.~N.\ , P.~S.\ and G.~S.\ gratefully acknowledge financial support by the
University Paris 7 Denis Diderot, where part of this work was
done. S.~N.\ and G.~S.\ also received support through US-NSF grant
no.\ DMS-0245210, P.~S.\ received support from the DFG.
%-------------------------------------------------------------%

%-------------------------------------------------------------%
\end{document}